\documentclass[11pt]{article}
\usepackage{amssymb,amsmath,graphicx}
\usepackage{amsfonts,setspace}
\usepackage{caption,natbib}
\usepackage[caption=false]{subfig}
\usepackage{epsfig,latexsym,graphicx}
\newtheorem{theorem}{Theorem}[section]

\newtheorem{remark}{Remark}[section]

\begin{document}

\title{Robust Parametric Inference for Finite Markov Chains 
}

\author{Abhik Ghosh 
\\
Indian Statistical Institute, Kolkata, India  
\\
{\it abhik.ghosh@isical.ac.in
}}
\maketitle

\begin{abstract}
We consider the problem of statistical inference in a parametric finite Markov chain model
and develop a robust  estimator of the parameters defining the transition probabilities 
via minimization of a suitable (empirical) version of the popular density power divergence.
Based on a long sequence of observations from a first-order stationary Markov chain, 
we have defined the minimum density power divergence estimator (MDPDE) of the underlying parameter 
and rigorously derived its asymptotic and robustness properties under appropriate conditions. 
Performance of the MDPDEs is illustrated theoretically as well as empirically for some common examples 
of finite Markov chain models. Its applications in robust testing of statistical hypotheses are also discussed
along with (parametric) comparison of two Markov chain sequences.
Several directions for extending the MDPDE and related inference 
are also briefly discussed for multiple sequences of Markov chains,
higher order Markov chains and non-stationary Markov chains with time-dependent transition probabilities. 
Finally, our proposal is applied to analyze corporate credit rating migration data of three international markets.
\end{abstract}

\bigskip

\section{Introduction}\label{SEC:intro}

Finite Markov chain models and their probabilistic characteristics are widely used to explain the behavior of physical systems or phenomenons;
such understanding of physical mechanisms is further applied to answer important research questions in psychology, genetics, epidemiology
and also several types of social studies \citep{Iosifescu:2007}.
For these applications, it is important to estimate the probabilistic structure 
of the assumed Markov chain model based on the sample data from the underlying physical process(es).

Consider one long, unbroken sequence $\mathcal{X}_T = \{X_0, X_1, \ldots, X_T\}$ 
of $(T+1)$ observations from a stationary Markov chain with finite state-space $S=\left\{1, 2, \ldots, K \right\}$
and transition probability matrix $\boldsymbol{\Pi}=(\pi_{ij})_{i,j=1, \ldots,K}$.
Note that, for each $i=1, \ldots, K$, the vector $\boldsymbol{\pi}_i=(\pi_{i1}, \ldots, \pi_{iK})$ is a probability measure on $S$;
let us denote all such probability vectors over $S$ by $\mathcal{P}_S$.
By stationarity,  the initial probability $\pi_{io} =  P(X_t=i)$ is independent of $t$ for each $i=1, \ldots, K$.
We assume that the Markov chain is ergodic (irreducible and aperiodic)
and consider the problem of making inference about these unknown probabilities $\pi_{ij}$s
and $\pi_{io}$s based on the observed sequence $\mathcal{X}_T$.
Assuming no further structure, their non-parametric (maximum likelihood) estimates are, respectively,  given by
\begin{eqnarray}
\widehat{\pi}_{ij} = \frac{\nu_{ij}}{\nu_{i+}},
~~\mbox{ and }~~
\widehat{\pi}_{io} = \frac{\nu_{i+}}{T},
~~~i,j=1, \ldots, K,
\label{EQ:Est_np}
\end{eqnarray}  
where $I(A)$ denotes the indicator function of the event $A$,
$\nu_{ij}=\sum_{t=0}^{T-1}I(X_t=i,X_{t+1}=j)$ and  $\nu_{i+}=\sum_{j=1}^K \nu_{ij}$ for $i,j=1, \ldots, K$.
The estimated transition probability matrix is then given by 
$\widehat{\boldsymbol{\Pi}}=(\widehat{\pi}_{ij})_{i,j=1, \ldots,K}$.
More details about these estimates and their asymptotic properties can be found 
in, e.g., \cite{Jones:2004}, \cite{Rajarshi:2014} and the references therein.  
%
However, in several applications in epidemiology, biology, Genomics, reliability studies, etc., 
we often model the transition probability matrix $\boldsymbol{\Pi}$ by a parametric model family 
of $K\times K$ transition matrices 
$\mathcal{F} = \left\{\boldsymbol{P}(\boldsymbol{\theta}) = (p_{ij}(\boldsymbol{\theta}))_{i,j=1, \ldots,K}
: \boldsymbol{\theta}\in\Theta\subseteq\mathbb{R}^d \right\}$, 
where $p_{ij}(\boldsymbol{\theta})$ are known functions depending on some unknown $d$-dimensional parameter vector 
$\boldsymbol{\theta}=(\theta_1, \ldots, \theta_d)^t\in \Theta$, the parameter space, 
and $\boldsymbol{P}_i(\boldsymbol{\theta}) = (p_{i1}(\boldsymbol{\theta}), \ldots, p_{iK}(\boldsymbol{\theta}))\in\mathcal{P}_S$  
for every $\boldsymbol{\theta}\in \Theta$, for each $i,j=1, \ldots, K$.
We assume, throughout this paper, that this model family $\mathcal{F}$ is identifiable in the sense 
$\boldsymbol{P}(\boldsymbol{\theta}_1) = \boldsymbol{P}(\boldsymbol{\theta}_2)$, for any two 
parameter values $\boldsymbol{\theta}_1, \boldsymbol{\theta}_2 \in \Theta$, must imply 
$\boldsymbol{\theta}_1 = \boldsymbol{\theta}_2$.
Then, any inference has to be performed based upon an estimate of $\boldsymbol{\theta}$ 
which is consistent and asymptotically normal under suitable regularity conditions.
The maximum likelihood estimator (MLE) is an immediate (optimum) candidate for this purpose,  
which were studied by \cite{Billingsley:1961} and is still the mostly used method of inference in a finite Markov chain.
Some modified likelihood based approach (e.g., PL, QL) are also developed for computational feasibility;
see \cite{Hjort/Varin:2008} and the references therein. 

Although asymptotically optimal, a well-known drawback of all these likelihood based inference
is their non-robustness against outliers or data contamination leading to erroneous insights.
Since outliers are not infrequent in several real life applications, 
a robust statistical procedure automatically taking care of the outliers is of great value 
to produce robust estimators and subsequent stable inference in such cases. 
An alternative to the MLE based on the minimum distance approach was discussed by \cite{Menendez/etc:1999} 
using disparity measures, but they did not discuss the issue of robustness.
In general the robustness concern in the present context of finite Markov chains is not well-studied,
although several works are available in the literature that developed robust procedures under general Markov (chains) models 
or important time series models as its special case; see, among many others, 
\cite{Birge:1983,Kunsch:1984,Martin/Yohai:1986,Lee/Lee:2010,Lee/Song:2013,Kang/Lee:2014,Bertail/etc:2018}.

Here, we will develop a robust methodology for 
parameter estimation and associated inference, particularly for the finite Markov chain models with a general parametric structure. 
As a way to solve the robustness issue, here, we consider the popular minimum distance approach 
based on the density power divergence (DPD) measure that was originally introduced by \cite{Basu/etc:1998} for IID data.
The DPD measure is a one-parameter generalization of the  Kullback-Leibler divergence (KLD);
for any two densities $g$ and $f$, with respect to some common dominating measure $\mu$,
the DPD measure is defined in terms of a tuning parameter $\alpha\geq 0$ as
\begin{eqnarray}\label{EQ:dpd}
d_\alpha(g,f) &=& \displaystyle \int  \left[f^{1+\alpha} - \left(1 + \frac{1}{\alpha}\right)  f^\alpha g + 
\frac{1}{\alpha} g^{1+\alpha}\right]d\mu,~~~\alpha\geq 0,\\
d_0(g,f) &=& \lim\limits_{\alpha\rightarrow 0} d_\alpha(g,f) = \int g \log(g/f)d\mu.
\end{eqnarray}
Note that, $d_0(g,f)$ is the KLD measure, and $d_1$ is the squared $L_2$ distance.
Since the MLE can be thought of as a minimizer of the KLD measure between the data and the model,
a generalized estimator can be obtained by minimizing the corresponding  DPD measure for any given $\alpha>0$.
The resulting minimum DPD estimator (MDPDE) has recently become popular due to the simplicity in its construction and computation
along with high robustness properties; 
it is also highly efficient with the tuning parameter $\alpha$ controlling  
the trade-offs between its efficiency and robustness \cite[see, e.g.,][]{Basu/etc:2011}. 
This approach, based on the MDPDE, is applied successfully in several disciplines and contexts
to produce robust insights against possible data contamination; 
see, e.g., \cite{Basu/etc:2006,Basu/etc:2018,Ghosh/Basu:2013,Ghosh/Basu:2018,Ghosh/etc:2016,Ghosh/etc:2018}, 
among many more. However, the MDPDE has not yet been developed for general set-ups of dependent observations 
except for  a few special cases like time series and diffusion processes in scattered attempts
\cite[e.g.,][and related works]{Lee/Lee:2010,Lee/Song:2013,Kang/Lee:2014}.   
In particular, advantages of the MDPDEs are yet to be investigated for finite Markov chains and other general stochastic processes.

In this paper, we study the MDPDEs for  the general class of finite Markov chain models 
as a robust generalization of the MLE and use it for further robust inference.
Although the implementation of the MDPDE in this context is somewhat straightforward, due to the dependence between observations, 
derivation of its theoretical properties needs non-trivial extensions of the existing results.
Such new asymptotic results for the MDPDE, defined under the finite Markov chain models, 
are major contributions of the present paper, along with several illustrative examples,
their use in robust inference (both confidence interval and hypothesis testing) and several important extensions.

We start by defining the MDPDE as a minimizer of  an appropriate (generalized) total discrepancy measure in terms of the DPD 
between rows of the empirical estimate $\widehat{\boldsymbol{\Pi}}$ and the model transition matrix  $\boldsymbol{P}(\boldsymbol{\theta})$,
and then derive its asymptotic and robustness properties. 
In particular, we prove the consistency and asymptotic normality of the MDPDE as $T\rightarrow\infty$ 
with suitable new arguments involving the Markov chain theory. 
The performance of the MDPDEs is illustrated through several interesting examples of finite Markov chains.
The asymptotic relative efficiency (ARE) of the MDPDEs (defined as the ratio of its asymptotic variance over that of the MLE) 
are used to study the effect of tuning parameter
and finite sample simulation studies are performed to justify the robustness benefits of the MDPDEs.
These illustrations indeed show, for the first time, the usefulness of the robust MDPDE for finite Markov chain models.

Further, we describe the application of these MDPDEs in performing statistical testing of general composite hypotheses.
Besides the derivation of general asymptotic properties of the MDPDE based Wald-type tests, 
an example of testing for the Bernoulli-Laplace diffusion model against a suitable parametric family of alternatives is discussed.
The MDPDE based testing procedure is also developed for (parametrically) comparing two Markov chain sequences.

Finally, we discuss important extensions of the MDPDE for a few complex finite Markov chain model set-ups.
These include the case of multiple sequence of observations obtained from the same finite Markov chain model,
where the asymptotics of the MDPDE are discussed for both the cases of diverging sequence length (with finite number of sequences)
and diverging number of observed sequences (of finite length each). 
Definitions of the MDPDE are also extended for higher-order Markov chains and those 
with time-dependent (non-stationary) transition probabilities. 

\noindent 
Due to page restrictions, all proofs and some illustrations are presented in the Online Supplement.

\section{Robust Estimation for A Finite Markov Chain}
\label{SEC:MDPDE}

\subsection{The Minimum Density Power Divergence Estimator}
\label{SEC:MDPDE_Est}

Let us consider the set-up and notation of Section \ref{SEC:intro}.
The widely popular MLE of $\boldsymbol{\theta}$ is defined as the maximizer of the likelihood function 
\vspace*{-0.3cm}
$$
L_T(\boldsymbol{\theta}) = p_{X_0X_1}(\boldsymbol{\theta})p_{X_1X_2}(\boldsymbol{\theta})\cdots p_{X_{T-1}X_T}(\boldsymbol{\theta})
=\prod_{i,j=1}^{K} p_{ij}(\boldsymbol{\theta})^{\nu_{ij}}.
$$
Then, the log-likelihood function is given by (See Online Supplement, Section A.1, for derivation)
\begin{eqnarray}
\frac{1}{T}\log L_T(\boldsymbol{\theta}) = -  \sum_{i=1}^K \widehat{\pi}_{io} \sum_{j=1}^K \widehat{\pi}_{ij} \log \frac{\widehat{\pi}_{ij}}{p_{ij}(\boldsymbol{\theta})} + \sum_{i=1}^K \widehat{\pi}_{io} \sum_{j=1}^K \widehat{\pi}_{ij} \log\widehat{\pi}_{ij}, 
\label{EQ:log-lik}
\end{eqnarray}
where the first term is a weighted average of KLD measure between the estimated probability vector 
$\widehat{\boldsymbol{\Pi}}_i=(\widehat{\pi}_{i1}, \ldots, \widehat{\pi}_{iK})$ and 
the model probability vector 
$\boldsymbol{P}_i(\boldsymbol{\theta}) = (p_{i1}(\boldsymbol{\theta}), \ldots, p_{iK}(\boldsymbol{\theta}))$ 
over different $i=1, \ldots, K$.  
Hence, the MLE can be obtained by minimizing such total KLD measure.
Since DPD is a generalization of the KLD measure at $\alpha>0$, in view of (\ref{EQ:log-lik}), 
we can define the MDPDE at any $\alpha> 0$ as the minimizer of a similarly defined total  DPD measure given by
$$
\sum_{i=1}^K \widehat{\pi}_{io} 
d_\alpha(\widehat{\boldsymbol{\Pi}}_i,\boldsymbol{P}_i(\boldsymbol{\theta}))
= \sum_{i=1}^K \widehat{\pi}_{io} \sum_{j=1}^K \left\{p_{ij}(\boldsymbol{\theta})^{1+\alpha}
- \left(1 + \frac{1}{\alpha}\right)  p_{ij}(\boldsymbol{\theta})^\alpha \widehat{\pi}_{ij} + 
\frac{1}{\alpha} \widehat{\pi}_{ij}^{1+\alpha}\right\},
$$
with respect to $\boldsymbol{\theta}\in \Theta$.
Since the last term within the bracket in the above equation does not depend on $\boldsymbol{\theta}$,
the MDPDE can be obtained by minimizing, in $\boldsymbol{\theta}\in \Theta$, the simpler objective function
\begin{eqnarray}
H_{T,\alpha}(\boldsymbol{\theta}) = (1+\alpha)^{-1}\boldsymbol{1}_K^t\boldsymbol{\Delta}(\widehat{\boldsymbol{\Pi}})
\boldsymbol{V}_\alpha(\widehat{\boldsymbol{\Pi}}, \boldsymbol{\theta})\boldsymbol{1}_K, 
\label{EQ:MDPDE_objFunc}
\end{eqnarray}
where $\boldsymbol{1}_K$ denotes a $K$-vector having each entry one, 
$\boldsymbol{\Delta}(\boldsymbol{\Pi}) = \mbox{Diag}\left\{ \pi_{io} ~:~ i=1, \ldots, K \right\}$ 
and $\boldsymbol{V}_\alpha({\boldsymbol{\Pi}}, \boldsymbol{\theta})$ is a $K\times K$ matrix having $(i,j)$-th  element as 
 $\left\{p_{ij}(\boldsymbol{\theta})^{1+\alpha} - \left(1 + \frac{1}{\alpha}\right)  
p_{ij}(\boldsymbol{\theta})^\alpha {\pi}_{ij}\right\}$.
Under the assumption of differentiability of $p_{ij}(\boldsymbol{\theta})$ in $\boldsymbol{\theta}$,
we can obtain the estimating equations of the MDPDE at any $\alpha>0$ as given by 
\vspace*{-0.3cm}
\begin{eqnarray}
\boldsymbol{U}_{T,\alpha}(\boldsymbol{\theta}) := \sum_{i=1}^K \widehat{\pi}_{io} \sum_{j=1}^K 
\boldsymbol{\psi}_{ij}(\boldsymbol{\theta})
\left(p_{ij}(\boldsymbol{\theta})- \widehat{\pi}_{ij}\right)p_{ij}(\boldsymbol{\theta})^{\alpha}  = \boldsymbol{0}_d,
\label{EQ:MDPDE_EstEq}
\end{eqnarray}
where $\boldsymbol{\psi}_{ij}(\boldsymbol{\theta}) 
= \frac{\partial}{\partial\boldsymbol{\theta}}\log p_{ij}(\boldsymbol{\theta})$ and $\boldsymbol{0}_d$ 
denotes a $d$-vector having all entries zero.
Note that, at $\alpha = 0$, the MDPDE estimating equation (\ref{EQ:MDPDE_EstEq})
coincides with the score equation corresponding to the MLE.
Thus, the MDPDE coincides with the MLE at $\alpha=0$ and provides its robust generalization at $\alpha>0$.
We next show that the MDPDE estimating equations are indeed unbiased at the model. 

In this regard, we define the statistical functional, say $\boldsymbol{F}_\alpha(\boldsymbol{\Pi})$, 
corresponding to the MDPDE with tuning parameter $\alpha\geq 0$, at any general (true) transition matrix $\boldsymbol{\Pi}$,
as the minimizer of 
$\sum_{i=1}^K {\pi}_{io}  d_\alpha({\boldsymbol{\Pi}}_i,\boldsymbol{P}_i(\boldsymbol{\theta}))$
with respect to $\boldsymbol{\theta}\in\Theta$, where $\boldsymbol{\Pi}_i$ denote the $i$-th row of $\boldsymbol{\Pi}$
and $\pi_{io}$s are the true initial probabilities depending on $\boldsymbol{\Pi}$. 
In consistence with the MDPDE objective function in (\ref{EQ:MDPDE_objFunc}), 
we can write $\boldsymbol{F}_\alpha(\boldsymbol{\Pi}) = \arg\min_{\boldsymbol{\theta}} H_{\alpha}(\boldsymbol{\Pi},\boldsymbol{\theta})$,
where $H_{\alpha}(\boldsymbol{\Pi},\boldsymbol{\theta}) = (1+\alpha)^{-1}\boldsymbol{1}_K^t\boldsymbol{\Delta}(\boldsymbol{\Pi})
\boldsymbol{V}_\alpha({\boldsymbol{\Pi}}, \boldsymbol{\theta})\boldsymbol{1}_K$.
The corresponding estimating equation for the MDPDE functional $\boldsymbol{F}_\alpha(\boldsymbol{\Pi})$  has the form 
\begin{eqnarray}
\boldsymbol{U}_{\alpha}(\boldsymbol{\Pi}, \boldsymbol{P}(\boldsymbol{\theta})) 
:= \sum_{i=1}^K {\pi}_{io} \sum_{j=1}^K \boldsymbol{\psi}_{ij}(\boldsymbol{\theta})
\left(p_{ij}(\boldsymbol{\theta})- {\pi}_{ij}\right)p_{ij}(\boldsymbol{\theta})^{\alpha}  = \boldsymbol{0}_d.
\label{EQ:MDPDEF_EstEq}
\end{eqnarray}
Note that, clearly $H_{\alpha}(\widehat{\boldsymbol{\Pi}}, \boldsymbol{\theta}) 
=H_{T,\alpha}(\boldsymbol{\theta})$ 
and $\boldsymbol{U}_{\alpha}(\widehat{\boldsymbol{\Pi}}, \boldsymbol{P}(\boldsymbol{\theta}))
=\boldsymbol{U}_{T,\alpha}(\boldsymbol{\theta})$
which implies $\boldsymbol{F}_\alpha(\widehat{\boldsymbol{\Pi}})$ is indeed the proposed MDPDE.
Further,  if the model is correctly specified with the true transition matrix being 
$\boldsymbol{P}(\boldsymbol{\theta}_0)$  for some $\boldsymbol{\theta}_0\in \Theta$, 
then  the estimating equation
$\boldsymbol{U}_{\alpha}(\boldsymbol{P}(\boldsymbol{\theta}_0), \boldsymbol{P}(\boldsymbol{\theta})) = \boldsymbol{0}_d$ 
has a solution at $\boldsymbol{P}(\boldsymbol{\theta}) =\boldsymbol{P}(\boldsymbol{\theta}_0)$.
Under the assumption of identifiability of our model family $\mathcal{F}$, 
it further implies $\boldsymbol{\theta}=\boldsymbol{\theta}_0$
and hence  $\boldsymbol{F}_\alpha(\boldsymbol{P}(\boldsymbol{\theta}_0))=\boldsymbol{\theta}_0$,
i.e., the MDPDE functional  $\boldsymbol{F}_\alpha$ is Fisher consistent at the model family $\mathcal{F}$
\cite[see, e.g.,][for definition of Fisher Consistency]{Basu/etc:2011}.
When the true transition matrix $\boldsymbol{\Pi}$ does not belong to the model family $\mathcal{F}$,
we will denote the corresponding MDPDE functional $\boldsymbol{\theta}_\pi = \boldsymbol{F}_\alpha(\boldsymbol{\Pi})$ 
as the `best fitting parameter" value (in the DPD sense)
and we will show that the corresponding MDPDE is also 
asymptotically consistent for this $\boldsymbol{\theta}_\pi$. 

Some illustrative examples of the MDPDE under finite Markov chain models are provided in Section \ref{SEC:Ex}.
Another interesting example, where the MDPDE equals to the MLE for every $\alpha\geq 0$, 
is presented in the Online Supplement (Section B).

\subsection{Asymptotic Properties}
\label{SEC:MDPDE_asymp}

In order to derive the asymptotic properties of the proposed MDPDE under the finite Markov chain models,
we first assume the following regularity conditions on the model transition probabilities.

\begin{itemize}
	\item[(A1)] For each $\boldsymbol{\theta}\in\Theta$, the model transition probability matrix 
	$\boldsymbol{P}(\boldsymbol{\theta})$ has the same sets of zero elements, i.e., 
	the set $C=\left\{(i,j): p_{ij}(\boldsymbol{\theta})>0 \right\}$ is independent of $\boldsymbol{\theta}$.
	Put $c=|C|$.\\
	Additionally, $C$ is regular in the sense that any Markov chain with transition probabilities $\pi_{ij}$ satisfying 
	``$\pi_{ij}>0$ if and only if $(i,j)\in C$" is irreducible \cite[see, e.g.,][]{Iosifescu:2007}.
	
	\item[(A2)] For every $(i,j)\in C$, the function $p_{ij}(\boldsymbol{\theta})$ 
	is twice continuously differentiable in $\boldsymbol{\theta}\in\Theta$.
	
	\item[(A3)] The $c\times d$ matrix $\boldsymbol{J}(\boldsymbol{\theta})=(J_{ij,u})_{(i,j)\in C, u=1, \ldots, d}$
	has rank $d$, for all $\boldsymbol{\theta}\in \Theta$, where 
	$J_{ij,u}=\frac{\partial p_{ij}(\boldsymbol{\theta})}{\partial\theta_u}.$
\end{itemize}
Based on (A1), for any $K\times K$ transition matrix $\boldsymbol{\Pi}\in \mathcal{P}_S^K$, we define 
the $c$-vector  $\boldsymbol{\Pi}_C$ having elements $\pi_{ij}$ only for $(i,j)\in C$
(the elements are stacked row-wise in our convention)
and denote the set of all such vectors as 
$\Im_C = \left\{\boldsymbol{\Pi}_C : \boldsymbol{\Pi} \in \mathcal{P}_S^K \right\}$.
Then, in view of Theorem 3.1 of  \cite{Billingsley:1961}, for a stationary and ergodic finite Markov chain 
having true transition matrix $\boldsymbol{\Pi}$, we have 
\begin{eqnarray}
\boldsymbol{\eta} := \sqrt{T}\left(\widehat{\boldsymbol{\Pi}}_{C} - \boldsymbol{\Pi}_C\right)
\mathop{\rightarrow}^\mathcal{D} \mathcal{N}_c\left(\boldsymbol{0}_c, \boldsymbol{\Lambda}(\boldsymbol{\Pi})\right),
~~~~\mbox{as } T\rightarrow\infty,
\label{EQ:CLT}
\end{eqnarray}
where $\widehat{\boldsymbol{\Pi}}=(\widehat{\pi}_{ij})$ from (\ref{EQ:Est_np}) and 
$\boldsymbol{\Lambda}(\boldsymbol{\Pi})=(\lambda_{ij,kl})_{(i,j), (k,l)\in C}$ is a $c\times c$ matrix 
having entries 
$
\lambda_{ij,kl} = \delta_{ik}\left(\delta_{jl}\pi_{ij} - \pi_{ij}\pi_{il}\right)/\pi_{io}.
$
The rate of convergence in (\ref{EQ:CLT}) is uniform in a neighborhood of $\boldsymbol{\Pi}$ 
and also  $\widehat{\boldsymbol{\Pi}}_{C} \rightarrow \boldsymbol{\Pi}_C$ almost surely (a.s.) as $T\rightarrow\infty$
\citep{Lifshits:1979,Sirazhdinov/Formanov:1984}. 

We now define a few matrices which are required for our asymptotic derivations.
For any $\boldsymbol{\Pi}\in \mathcal{P}_S^K$ satisfying (A1) and any $\boldsymbol{\theta}\in\Theta$, 
define the $c\times c$ matrix
$
\boldsymbol{B}_\alpha(\boldsymbol{\Pi}, \boldsymbol{\theta})=
\mbox{Diag}\left\{\frac{p_{ij}(\boldsymbol{\theta})^{1-\alpha}}{\pi_{io}} : (i,j)\in C \right\}.
$
Also, let us define two $d\times d$ matrices, which are non-singular by Assumption (A3), as given by 
$\boldsymbol{\Omega}_\alpha(\boldsymbol{\Pi}, \boldsymbol{\theta}) = 
\boldsymbol{J}(\boldsymbol{\theta})^t\boldsymbol{B}_\alpha(\boldsymbol{\Pi}, \boldsymbol{\theta})^{-1}
\boldsymbol{\Lambda}(\boldsymbol{\Pi})
\boldsymbol{B}_\alpha(\boldsymbol{\Pi}, \boldsymbol{\theta})^{-1}\boldsymbol{J}(\boldsymbol{\theta})
$
and 
\begin{eqnarray}
\boldsymbol{\Psi}_\alpha(\boldsymbol{\Pi}, \boldsymbol{\theta}) &=& \boldsymbol{J}(\boldsymbol{\theta})^t\boldsymbol{B}_\alpha(\boldsymbol{\Pi}, \boldsymbol{\theta})^{-1}
\boldsymbol{J}(\boldsymbol{\theta})
\nonumber\\
&& ~~~~+ \sum_{(i,j)\in C} \pi_{io}p_{ij}(\boldsymbol{\theta})^\alpha
\left[\alpha\boldsymbol{\psi}_{ij}(\boldsymbol{\theta})^T\boldsymbol{\psi}_{ij}(\boldsymbol{\theta})
+ \frac{\partial\boldsymbol{\psi}_{ij}(\boldsymbol{\theta})}{\partial\boldsymbol{\theta}}\right]
\left(p_{ij}(\boldsymbol{\theta}) - \pi_{ij}\right),
\end{eqnarray}

Now, let us first restrict ourselves to the cases where the assumed parametric model family is correctly specified 
and hence  the true transition probability matrix $\boldsymbol{\Pi}$ belongs to the model family,
i.e., $\boldsymbol{\Pi} =\boldsymbol{P}(\boldsymbol{\theta}_0)$ for some $\boldsymbol{\theta}_0\in \Theta$.
For simplicity, we put $\boldsymbol{P}^o=\boldsymbol{P}(\boldsymbol{\theta}_0)$.
Note that, in such cases we have, 
for any $\boldsymbol{\theta}\in\Theta$ (including $\boldsymbol{\theta}_0$), 
$
 \boldsymbol{\Psi}_\alpha(\boldsymbol{P}(\boldsymbol{\theta}), \boldsymbol{\theta}) = \boldsymbol{J}(\boldsymbol{\theta})^t\boldsymbol{B}_\alpha(\boldsymbol{\Pi}, \boldsymbol{\theta})^{-1}
 \boldsymbol{J}(\boldsymbol{\theta}).
$
From now on, we will use the notation $p_{io}(\boldsymbol{\theta}) := \pi_{io}$, the initial probabilities, 
when $\boldsymbol{\Pi}=\boldsymbol{P}(\boldsymbol{\theta})$
and assume that $\boldsymbol{P}_{io}(\boldsymbol{\theta})
=(p_{1o}(\boldsymbol{\theta}), \ldots, p_{Ko}(\boldsymbol{\theta}))\in \mathcal{P}_S$ for all $\boldsymbol{\theta}\in\Theta$.
Then, using the result (\ref{EQ:CLT}) with $\boldsymbol{\Pi}=\boldsymbol{P}^o$ and extending the arguments from \cite{Menendez/etc:1999}, 
we prove the asymptotic consistency of the MDPDEs at the model 
which is  presented in the following theorem.

\begin{theorem}
Consider a finite Markov chain that is stationary and ergodic having true transition matrix 
$\boldsymbol{P}^o=\boldsymbol{P}(\boldsymbol{\theta}_0)\in\mathcal{P}_S^K$ for  $\boldsymbol{\theta}_0\in\Theta$.
Fix $\alpha\geq 0$. 
Then, under Assumptions (A1)--(A3), we have the following results. 
\begin{itemize}
	\item[(i)] There exists a solution $\widehat{\boldsymbol{\theta}}_{\alpha}$ (MDPDE) 
	to the estimating equation (\ref{EQ:MDPDE_EstEq}) which is unique a.s. in a neighborhood of $\boldsymbol{\theta}_0$ 
	and satisfies the relation
\begin{eqnarray}
\sqrt{T}\left(\widehat{\boldsymbol{\theta}}_{\alpha} - \boldsymbol{\theta}_0\right)
= \boldsymbol{\Psi}_\alpha(\boldsymbol{P}^o, \boldsymbol{\theta}_0)^{-1}
\boldsymbol{J}(\boldsymbol{\theta}_0)^t\boldsymbol{B}_\alpha(\boldsymbol{P}^o, \boldsymbol{\theta}_0)^{-1}\boldsymbol{\eta}
+o_P(1), ~~~~\mbox{as } T\rightarrow\infty.
\label{EQ:MDPDE_asymEx}
\end{eqnarray} 
	\item[(ii)] The MDPDE $\widehat{\boldsymbol{\theta}}_{\alpha}$ is consistent for $\boldsymbol{\theta}_0$ 
	and also asymptotically normal with 
		\begin{eqnarray}
	\sqrt{T}\left(\widehat{\boldsymbol{\theta}}_{\alpha} - \boldsymbol{\theta}_0\right)
	\mathop{\rightarrow}^\mathcal{D} 
	\mathcal{N}_d\left(\boldsymbol{0}_d, \boldsymbol{\Sigma}_\alpha(\boldsymbol{P}^o, \boldsymbol{\theta}_0)\right),
	~~~~\mbox{as } T\rightarrow\infty,
	\label{EQ:MDPDE_AN}
	\end{eqnarray} 
	where $\boldsymbol{\Sigma}_\alpha(\boldsymbol{\Pi}, \boldsymbol{\theta})
	= \boldsymbol{\Psi}_\alpha(\boldsymbol{\Pi}, \boldsymbol{\theta})^{-1}
	\boldsymbol{\Omega}_\alpha(\boldsymbol{\Pi}, \boldsymbol{\theta})
	\boldsymbol{\Psi}_\alpha(\boldsymbol{\Pi}, \boldsymbol{\theta})^{-1}$.
\end{itemize}
\label{THM:MDPDE_conv}
\end{theorem}
Theorem \ref{THM:MDPDE_conv} can also be extended under  model misspecifications, 
where the true transition matrix, say $\boldsymbol{\Pi}^o$, 
does not belong to the assumed model family. In this case,
we can talk about the consistency only at the ``best fitting parameter value"
$\boldsymbol{\theta}_\pi =\boldsymbol{F}_\alpha(\boldsymbol{\Pi}^o)$ defined in Section \ref{SEC:MDPDE_Est}.
Then, the conclusions of Theorem \ref{THM:MDPDE_conv} still hold with slight modifications
as given in the next theorem.

\begin{theorem}
	Consider a finite Markov chain that is stationary and ergodic having true transition matrix 
	$\boldsymbol{\Pi}^o$, which does not necessarily belongs to the model family $\mathcal{F}$,
	and fix an $\alpha\geq 0$. Let 	$\boldsymbol{\theta}_\pi =\boldsymbol{F}_\alpha(\boldsymbol{\Pi}^o)$
	denote the ``best fitting parameter value" in the DPD sense.
	Then, under Assumptions (A1)--(A3), we have the following results. 
	\begin{itemize}
		\item[(i)] There exists a solution $\widehat{\boldsymbol{\theta}}_{\alpha}$ (MDPDE) 
		to the estimating equation (\ref{EQ:MDPDE_EstEq}) which is unique a.s. in a neighborhood of $\boldsymbol{\theta}_\pi$ 
		and satisfies the relation
		\begin{eqnarray}
		\sqrt{T}\left(\widehat{\boldsymbol{\theta}}_{\alpha} - \boldsymbol{\theta}_\pi\right)
		= \boldsymbol{\Psi}_\alpha(\boldsymbol{\Pi}^o, \boldsymbol{\theta}_\pi)^{-1}
		\boldsymbol{J}(\boldsymbol{\theta}_\pi)^t\boldsymbol{B}_\alpha(\boldsymbol{\Pi}^o, \boldsymbol{\theta}_\pi)^{-1}\boldsymbol{\eta}
		+o_P(1), ~~~~\mbox{as } T\rightarrow\infty.
		\label{EQ:MDPDE_asymEx_Out}
		\end{eqnarray} 
		\item[(ii)] The MDPDE $\widehat{\boldsymbol{\theta}}_{\alpha}$ is consistent for $\boldsymbol{\theta}_\pi$ and 
		$\sqrt{T}\left(\widehat{\boldsymbol{\theta}}_{\alpha} - \boldsymbol{\theta}_\pi\right)
		\mathop{\rightarrow}^\mathcal{D} 
		\mathcal{N}_d\left(\boldsymbol{0}_d, \boldsymbol{\Sigma}_\alpha(\boldsymbol{\Pi}^o, \boldsymbol{\theta}_\pi)\right)$
		as $T\rightarrow\infty$.
	\end{itemize}
	\label{THM:MDPDE_conv_Out}
\end{theorem}

The proof of Theorem \ref{THM:MDPDE_conv_Out} is given in  the Online Supplement (Section A.2).
Then, Theorem \ref{THM:MDPDE_conv} follows from Theorem \ref{THM:MDPDE_conv_Out} as a special case 
by substituting $\boldsymbol{\Pi}^o$ and $\boldsymbol{\theta}_\pi$ by $\boldsymbol{P}^0$ and $\boldsymbol{\theta}_0$, respectively. 
 
\begin{remark}[The special case $\alpha=0$]~~~\\
	We have already argued that the MDPDE is a generalization of the classical MLE and,
	in fact, coincides with the MLE at $\alpha=0$. 
We can also find out the asymptotic distribution of the MLE, at the model, as a special case of Theorem \ref{THM:MDPDE_conv} at $\alpha=0$. 
	In particular, some algebra lead us to 
	$\boldsymbol{\Psi}_0(\boldsymbol{P}^o, \boldsymbol{\theta}_0) =
	\boldsymbol{\Omega}_0(\boldsymbol{P}^o, \boldsymbol{\theta}_0)$.
	Then the asymptotic variance of ($\sqrt{T}$ times) MLE  turns out to be 
	$\boldsymbol{\Psi}_0(\boldsymbol{P}^o, \boldsymbol{\theta}_0)^{-1}$.
	This coincides with the usual maximum likelihood theory, since 
	$\boldsymbol{\Psi}_0(\boldsymbol{P}^o, \boldsymbol{\theta}_0)$ is indeed the Fisher information matrix 
	of our model. Further, interestingly, the minimum disparity estimators, 
	discussed in \cite{Menendez/etc:1999}, also have the same asymptotic distribution as that of the MDPDE at $\alpha=0$.
	Further, putting $\alpha=0$, Theorem \ref{THM:MDPDE_conv_Out} provides an additional result in this context, 
	yielding the asymptotic distribution of the MLE under model misspecification.
\end{remark}

\subsection{Robustness of the MDPDE}
\label{SEC:IF_MDPDE}

Suppose that the data are observed from a stationary ergodic finite Markov chain  having true transition matrix 
$\boldsymbol{\Pi}^o$, which does not necessarily belong to the model family $\mathcal{F}$.
Consider a contaminated transition matrix 
$\boldsymbol{\Pi}_\epsilon = (1-\epsilon)\boldsymbol{\Pi}^o + \epsilon \boldsymbol{D}_{\boldsymbol{t}}$
where $\epsilon\in[0,1]$ denote the contamination proportion, 
$\boldsymbol{t}=(t_1, \ldots, t_K) \in S^K$ is the contamination point  
and  $\boldsymbol{D}_{\boldsymbol{t}}$ has entry one at $(i, t_i)$-th position 
for all $i=1, \ldots, K$  and zero in all other positions. 
These leads to contaminated probability vector for each row of the transition matrix.
We then define the influence function (IF) of the MDPDE functional $\boldsymbol{F}_\alpha(\boldsymbol{\Pi})$  at a fixed $\alpha\geq 0$,
under the present finite Markov model set-up, as
$$
IF(\boldsymbol{t}; \boldsymbol{F}_\alpha, \boldsymbol{\Pi}^o)
= \lim_{\epsilon\downarrow 0} \frac{\boldsymbol{F}_\alpha(\boldsymbol{\Pi}_\epsilon) - \boldsymbol{F}_\alpha(\boldsymbol{\Pi}^o)}{\epsilon} 
= \frac{\partial}{\partial\epsilon}\boldsymbol{F}_\alpha(\boldsymbol{\Pi}_\epsilon)\bigg|_{\epsilon=0}.
$$ 
It is necessary to justify the above definition of IF in the present context with dependence data;
this is important in the view of the existing literature discussing that the IF of \cite{Hampel/etc:1986} 
is not necessarily correct for some dependent situations, e.g., time series \citep{Martin/Yohai:1986}.
Although the IF was initially defined for IID data in \cite{Hampel/etc:1986}, it has latter been extended for several other scenarios
of non-homogeneous data; see, e.g., \cite{Ghosh/Basu:2013,Ghosh/etc:2016}. 
Although we have dependent data following a Markov chain, in above, we have defined IF in terms of transition matrix of the whole data
instead of the dependent distributions of the individual observations $X_i$s, which allows us to avoid the associated complications. 
Since a transition matrix uniquely characterizes the distribution of the whole sample together, 
our definition of the IF, in the line of \cite{Hampel/etc:1986}, is perfectly justified. 
Additionally, our definition indeed measures the amount of (asymptotic) bias  of the MDPDE functional $\boldsymbol{F}_\alpha(\boldsymbol{\Pi})$ 
against infinitesimal contamination at ${\boldsymbol{t}}$;
this  is also in the same spirit as that of the general concept of IF.  
However, the contamination here is characterized by transition matrices leading 
to possible missclassifications within the state-spaces (via ${\boldsymbol{t}}$), 
rather than a distant outlying observation as the case was in its classical definition by \cite{Hampel/etc:1986}.

Now, in order to derive this IF in the present context, we note that $\boldsymbol{F}_\alpha(\boldsymbol{\Pi}_\epsilon)$ 
satisfies the estimating equation (\ref{EQ:MDPDEF_EstEq}) with $\boldsymbol{\Pi}$ replaced by $\boldsymbol{\Pi}_\epsilon$, 
i.e., 
$
\boldsymbol{U}_{\alpha}(\boldsymbol{\Pi}_\epsilon, \boldsymbol{P}(\boldsymbol{F}_\alpha(\boldsymbol{\Pi}_\epsilon))) 
=\boldsymbol{0}_d.
$
Differentiating it with respect to $\epsilon$ and evaluating at $\epsilon=0$, 
we can get the IF of the MDPDE functional as
presented in the following theorem. The proof is given in the Online Supplement (Section A.3).

\begin{theorem}
	Consider a finite Markov chain that is stationary and ergodic having true transition matrix $\boldsymbol{\Pi}^o$	
	and fix an $\alpha\geq 0$. 
	Let 	$\boldsymbol{\theta}_\pi =\boldsymbol{F}_\alpha(\boldsymbol{\Pi}^o)$
	denote the ``best fitting parameter value" in the DPD sense.
	Then, the influence function of the MDPDE functional $\boldsymbol{F}_\alpha$ is given by 
	\begin{eqnarray}
	IF(\boldsymbol{t}; \boldsymbol{F}_\alpha, \boldsymbol{\Pi}^o) 
\label{EQ:MDPDE_IF}
	&=& \boldsymbol{\Psi}_\alpha(\boldsymbol{\Pi}^o, \boldsymbol{\theta}_\pi) ^{-1}\sum_{i=1}^K {\pi}_{io}  
	\left[\sum_{j=1}^K \boldsymbol{\psi}_{ij}(\boldsymbol{\theta}_\pi)p_{ij}(\boldsymbol{\theta}_\pi)^{\alpha}{\pi}_{ij}^o  
	- \boldsymbol{\psi}_{it_i}(\boldsymbol{\theta}_\pi)p_{it_i}(\boldsymbol{\theta}_\pi)^{\alpha}
	\right].
	\end{eqnarray}
	The above formula can be further simplified at the model where 
	$\boldsymbol{\Pi}^o=\boldsymbol{P}(\boldsymbol{\theta}_0)$ for some $\boldsymbol{\theta}_0\in\Theta$. 
	\label{THM:MDPDE_IF}
\end{theorem}

The only term of the IF that depends on the contamination point is 
$\varphi_{\alpha}(t_i) = \boldsymbol{\psi}_{it_i}(\boldsymbol{\theta}_\pi)p_{it_i}(\boldsymbol{\theta}_\pi)^{\alpha}$;
but this term is bounded for each $i$ and every $\alpha\geq0$ because of the finite state space.
So, in this case, the boundedness of the IF is always evident and we cannot talk about the robustness 
by just looking at the bounded IF. Note that, the contamination considered here is not far away from the majority of the data,
rather it covers the missclassification within the state-spaces arising from small-scale errors.
So, using this IF, we can still talk about the robustness under such contaminations in relative terms (and not as an absolute measure),
i.e., compare two estimators by looking at the maximum value of the IF (in an appropriate norm) over all possible $\boldsymbol{t}$ 
(misclassification types); as this maximum becomes smaller the estimator becomes more robust. 
This measure of quantifying the extent of robustness, commonly known as the sensitivity measure, is formally defined as
$\gamma_\alpha(\boldsymbol{\Pi}^o) 
= \sup_{\boldsymbol{t}\in S^K }\left|\left|IF(\boldsymbol{t}; \boldsymbol{F}_\alpha, \boldsymbol{\Pi}^o) \right|\right|.$
For most common examples, this sensitivity  indeed decreases with increasing $\alpha>0$, indicating 
the gain in robustness by the MDPDEs at larger $\alpha>0$.

An alternative (global) measure of robustness, that can provide further insights about the (absolute) 
robustness in the present case of finite state-spaces, 
is the maxbias curve of the functional defined as its maximum asymptotic bias under all possible contaminated (misspecified) distributions
for a given contamination proportion. Although Theorem \ref{THM:MDPDE_conv_Out} provides some insights about the consistency of 
the MDPDE to the best fitting parameter (with respect to the DPD) even under such contaminated scenarios, 
derivation of the exact form of the maxbias curve is not straightforward. 
But, it would surely be interesting and important future work to study the maxbias curve for the MDPDE under different model set-ups 
including the present finite Markov chain models.

\subsection{Estimating Standard Error  and Confidence Interval }
\label{SEC:IF_CI}

In any application, it is important to report the standard error of the estimator used.
The same can also be done for the MDPDEs under the present set-up by using the asymptotic variance formulas 
given in Theorems \ref{THM:MDPDE_conv} and \ref{THM:MDPDE_conv_Out}.
Firstly, based on Theorem \ref{THM:MDPDE_conv}, a consistent  estimate of the asymptotic variance matrix of ($\sqrt{T}$ times) 
the  MDPDE can be obtained as $\boldsymbol{\Sigma}_\alpha(\boldsymbol{P}(\widehat{\boldsymbol{\theta}}_{\alpha}), 
\widehat{\boldsymbol{\theta}}_{\alpha})$;
note that this estimator is model specific.
Alternatively, based on Theorem \ref{THM:MDPDE_conv_Out}, another model-robust estimator of the asymptotic variance matrix is given by
$\boldsymbol{\Sigma}_\alpha(\widehat{\boldsymbol{\Pi}}, \widehat{\boldsymbol{\theta}}_{\alpha})$;
this can be shown to be a consistent variance estimator under standard regularity conditions.
This second estimator works better compared to the previous model specific variance estimator 
$\boldsymbol{\Sigma}_\alpha(\boldsymbol{P}(\widehat{\boldsymbol{\theta}}_{\alpha}), \widehat{\boldsymbol{\theta}}_{\alpha})$
under model misspecification, but the first one works better against outliers with respect to a fixed model.
Using either of these two variance estimators, and denoting it by 
$\widehat{\boldsymbol{\Sigma}}=\left(\widehat{\sigma}_{ij}\right)_{i,j=1, \ldots, d}$, 
the standard error of the $i$-th component of $\boldsymbol{\theta}$ is then obtained as $\widehat{s}_i=\sqrt{\widehat{\sigma}_{ii}/T}$ 
for each $i=1, \ldots, d$.

Once we estimate the standard error of the parameter estimate, its immediate use is to construct the confidence intervals. 
Utilizing the asymptotic normality of the MDPDEs, we can actually construct 
the corresponding robust asymptotic confidence interval of the underlying parameters. 
In particular, a robust asymptotically $100(1-\zeta)\%$ confidence interval for the $i$-th component ($\theta_i$) 
of the parameter vector $\boldsymbol{\theta}$ is given by 
$(\widehat{\theta}_i - \widehat{s}_iz_{\zeta/2}, \widehat{\theta}_i + \widehat{s}_iz_{\zeta/2})$, 
where $\widehat{s}_i$ is the estimated standard error of the MDPDE $\widehat{\theta}_i$ of $\theta_i$ as defined above
and $z_{\tau}$ is the $(1-\tau)$-th quantile of the standard normal distribution ($\tau=\zeta/2$ here). 
These confidence intervals can be combined over all $i$, along with the necessary Boferroni correction, 
to obtain a robust and asymptotic simultaneous confidence set (rectangle) for the whole parameter vector $\boldsymbol{\theta}$.
Alternatively, using the joint asymptotic distribution of the MDPDE of $\boldsymbol{\theta}$ and the variance matrix estimator  
$\widehat{\boldsymbol{\Sigma}}$, one can also construct, in a straightforward way, 
robust asymptotic simultaneous confidence ellipsoid for the whole parameter vector. 
Such asymptotic confidence  sets for any (differentiable) function of the parameters can also be constructed 
by inverting the MDPDE based Wald-type tests to be introduced  in Section \ref{SEC:Hyp_test}.

\section{Examples and Illustrations }
\label{SEC:Ex}

\subsection{Example 1: A Random Walk Type Model with Binomial Probabilities}

Let us consider an example of finite Markov chain, over the state-space $S=\{1, 2, \ldots, K \}$,
with reflecting barriers and Bin(2, $\theta$) distribution for moving from each internal position 
to its nearest (both sided) three positions. The corresponding transition matrix is then given by
\begin{eqnarray}
\boldsymbol{P}(\theta) = \begin{bmatrix}
\begin{array}{cccccccc}
0 		 & 1 		& 0 		& 0 	 & \cdots & 0  & 0 & 0 \\
(1-\theta)^2 & 2\theta(1-\theta) 		& \theta^2 	& 0 	 & \cdots & 0  & 0 & 0\\
0 		 & (1-\theta)^2 & 2\theta(1-\theta)	& \theta^2 & \cdots & 0  & 0 & 0 \\
: 		 & 		: 	& 	: 		&  :	 &  \ddots & :  & : \\
0 		 & 	0		& 	0		& 0		& \cdots & (1-\theta)^2 & 2\theta(1-\theta)	& \theta^2  \\
0 		 &    0		&  	0		&  0     & \cdots  & 0 & 1  & 0 \\
\end{array}
\end{bmatrix}.
\label{EX2:P}
\end{eqnarray}Such a model often arise in real-life applications, e.g., in genetics, with different values of $K$.
Here, the target parameter $\theta\in\Theta = [0,1]$ is scalar
and the Markov chain is stationary and ergodic with initial (stationary) probabilities 
$\{\pi_{io}=p_{io}(\theta) : i=1, 2, \ldots, K \}$, where 
\begin{eqnarray}
&&p_{1o}(\theta) = \frac{(1-\theta)^{2(K-1)}(1-2\theta)}{2\left[(1-\theta)^{2K-1} - \theta^{2K-1}\right]},
~~~p_{Ko}(\theta) = \frac{\theta^{2(K-1)}(1-2\theta)}{2\left[(1-\theta)^{2K-1} - \theta^{2K-1}\right]},
\nonumber\\
\mbox{and }&&p_{io}(\theta) =  \frac{\theta^{2(i-1)}}{(1-\theta)^{2i}}\frac{(1-\theta)^{2(K-1)}(1-2\theta)}{2\left[(1-\theta)^{2K-1} - \theta^{2K-1}\right]}, 
~~~i=2, \ldots, K-1.  
\nonumber
\end{eqnarray}
Further, Assumptions (A1)--(A3) also hold for $\boldsymbol{P}(\theta)$ in (\ref{EX2:P}) with 
$$
C=\{ (1,2); (i, i+1), (i, i), (i, i-1) \mbox{ for } i=2, 3, \ldots, K-1; (K, K-1) \},
$$ 
so that $c=3K-4$ and 
$$
\boldsymbol{J}(\theta) = \big[0, 2\theta, 2(1-2\theta), -2(1-\theta), 2\theta, 2(1-2\theta), -2(1-\theta), 
\ldots, 2\theta, 2(1-2\theta), -2(1-\theta), 0\big]^t.
$$ 

Now consider one long sequence $\mathcal{X}_T = \{X_0, X_1, \ldots, X_T\}$  observed from this given  Markov chain
based on which we wish to infer about the target parameter $\theta$.
One can easily verify that the MLE of $\theta$ is given by 
$$
\widehat{\theta}_0 = \frac{\sum_{i=2}^{K-1} [\nu_{i(i+1)} + \nu_{ii}/2]}{\sum_{i=2}^{K-1} [\nu_{i(i-1)}+\nu_{ii} + \nu_{i(i+1)}]} 
= \frac{\sum_{i=2}^{K-1} \widehat{\pi}_{io} [\widehat{\pi}_{i(i+1)} + \widehat{\pi}_{ii}/2]}{\sum_{i=2}^{K-1} \widehat{\pi}_{io}}.
$$

On the other hand, the proposed MDPDE of $\theta$ with tuning parameter $\alpha> 0$ can be obtained 
by solving the estimating equation (\ref{EQ:MDPDE_EstEq}), which simplifies for the present case as 
\begin{eqnarray}
&& \sum_{i=2}^{K-1}  \left[ \theta^{2\alpha-1} \nu_{i(i+1)}
+ 2^{\alpha-1}\theta^{\alpha-1}(1-\theta)^{\alpha-1}(1-2\theta) \nu_{ii}
-  (1-\theta)^{2\alpha-1} \nu_{i(i-1)}\right]
\nonumber\\
&& ~~~~~~~~~~= \left(\sum_{i=2}^{K-1} \nu_{i+}\right) \left[\theta^{2\alpha+1} + 2^\alpha\theta^\alpha(1-\theta)^\alpha(1-2\theta)
- (1-\theta)^{2\alpha+1}\right]. 
\label{EQ:Ex2_MDPDE_EstEq}
\end{eqnarray}
%
Next we derive the asymptotic distribution of the MDPDE at the model using Theorem \ref{THM:MDPDE_conv}.
The required assumptions clearly hold for this example and, through some algebra, we obtain
$\boldsymbol{\Psi}_\alpha(\boldsymbol{P}(\theta), {\theta}) 
= 4[1-p_{1o}(\theta) - p_{Ko}(\theta))]V_{1,\alpha}(\theta)$
and
$\boldsymbol{\Omega}_\alpha(\boldsymbol{P}(\theta), {\theta}) = 4[1-p_{1o}(\theta) - p_{Ko}(\theta))]V_{2,\alpha}(\theta)$,
where
\begin{eqnarray}
V_{1,\alpha}(\theta) &=& (1-\theta)^{2\alpha} + \theta^{2\alpha} 
+ 2^{\alpha-1}\theta^{\alpha-1}(1-\theta)^{\alpha-1}(1-2\theta)^2,
\label{EQ:V1}\\
V_{2,\alpha}(\theta) &=& (1-\theta)^{4\alpha}\theta(2-\theta) + \theta^{4\alpha}(1-\theta^2) + 2\theta^{2\alpha+1}(1-\theta)^{2\alpha+1}
\nonumber\\
&& ~ ~+ 2^{\alpha+1}\theta^{\alpha}(1-\theta)^{3\alpha+1}(1-2\theta) - 2^{\alpha+1}\theta^{3\alpha+1}(1-\theta)^{\alpha}(1-2\theta)
\nonumber\\
&& ~~~~+ 2^{2\alpha-1}\theta^{2\alpha-1}(1-\theta)^{2\alpha-1}(1-2\theta)^2(1-2\theta+2\theta^2).
\label{EQ:V2}
\end{eqnarray}
Then, the asymptotic variance of $\sqrt{T}~\widehat{\theta}_{\alpha}$ is given by 
\begin{eqnarray}
\Sigma_\alpha(\boldsymbol{P}(\theta), {\theta}) =
[1-p_{1o}(\theta) - p_{Ko}(\theta)]^{-1}\frac{V_{2,\alpha}(\theta)}{4 V_{1,\alpha}(\theta)^2}.
\label{EQ:Ex2_AsymVar}
\end{eqnarray}
It is easy to see that this ${\Sigma}_\alpha(\boldsymbol{P}(\theta), {\theta})$  is 
symmetric about $\theta=1/2$ for each $\alpha\geq 0$, i.e., 
${\Sigma}_\alpha(\boldsymbol{P}(\theta), {\theta}) = {\Sigma}_\alpha(\boldsymbol{P}(1-\theta), {1-\theta})$,
 and is independent of $\alpha$ at $\theta=1/2$
having value ${\Sigma}_\alpha(\boldsymbol{P}(1/2), 1/2) = \frac{3}{4}[1-p_{1o}(\theta) - p_{Ko}(\theta)]^{-1}$. 
At any other fixed parameter value $\theta\neq 1/2$, 
 ${\Sigma}_\alpha(\boldsymbol{P}(\theta), {\theta})$ is a strictly increasing function of $\alpha\geq0$.
 In particular, at $\alpha=0$, we have the least asymptotic variance for the  MLE ($\sqrt{T}~\widehat{\theta}_0$) as given by 
$$ 
{\Sigma}_0(\boldsymbol{P}(\theta), {\theta}) =
\frac{2\theta(1-\theta)(8\theta^4 - 16\theta^3 + 8\theta^2 + 1)}{[1-p_{1o}(\theta) - p_{Ko}(\theta)]}.
$$
So, the ARE of the proposed MDPDE at any fixed $\alpha$ can be obtained by comparing 
 ${\Sigma}_\alpha(\boldsymbol{P}(\theta), {\theta})$ with ${\Sigma}_0(\boldsymbol{P}(\theta), {\theta})$,
 which are reported in Table \ref{TAB:Ex2_ARE} for different parameter values.
Note that, the ARE clearly decreases as $\alpha$ increases
but the loss in efficiency is not quite significant at smaller values of $\alpha>0$.
With this small price, these MDPDEs gain significant robustness against data contamination as illustrated below through 
a simulation study.

\begin{table}[h]
	\centering
	\caption{ARE (in \%) of the MDPDEs for Example 1 at different values of $\alpha>0$ and $\theta\in(0,1)$}
	\begin{tabular}{l|rrrrrr} \hline
$\theta$ or $(1-\theta)$		& \multicolumn{6}{|c}{$\alpha$}\\
	&	0.1	&	0.2	&	0.3	&	0.5	&	0.7	&	1	\\\hline
0.05	&	99.1	&	97.5	&	96.0	&	94.2	&	93.8	&	94.4	\\
0.1	&	98.9	&	96.7	&	94.3	&	90.8	&	89.2	&	89.2	\\
0.15	&	98.9	&	96.5	&	93.7	&	88.8	&	85.9	&	84.4	\\
0.2	&	99.1	&	96.8	&	93.9	&	88.2	&	83.9	&	80.7	\\
0.25	&	99.3	&	97.4	&	94.7	&	88.8	&	83.6	&	78.4	\\
0.3	&	99.5	&	98.1	&	96.0	&	90.7	&	85.2	&	78.4	\\
0.35	&	99.7	&	98.8	&	97.4	&	93.5	&	88.7	&	81.4	\\
0.4	&	99.9	&	99.4	&	98.7	&	96.6	&	93.5	&	87.9	\\
0.45	&	100.0	&	99.9	&	99.7	&	99.0	&	98.1	&	96.0	\\
0.5	&	100	&	100	&	100	&	100	&	100	&	100	\\
	\hline
	\end{tabular}
	\label{TAB:Ex2_ARE}
\end{table} 

\begin{figure}[!b]
	\centering
	\subfloat[$T=50$]{
		\includegraphics[width=0.5\textwidth]{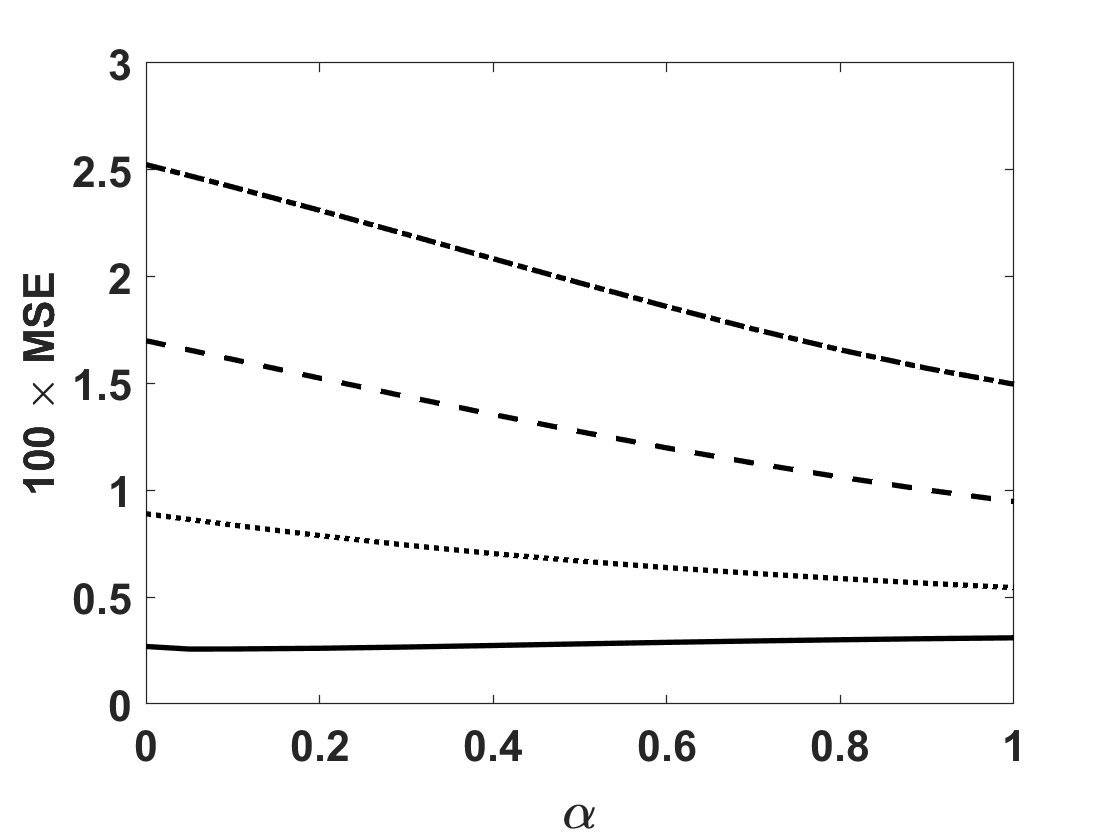}
		\label{FIG:Ex2_MSE_n50}}
	\subfloat[$T=100$]{
		\includegraphics[width=0.5\textwidth]{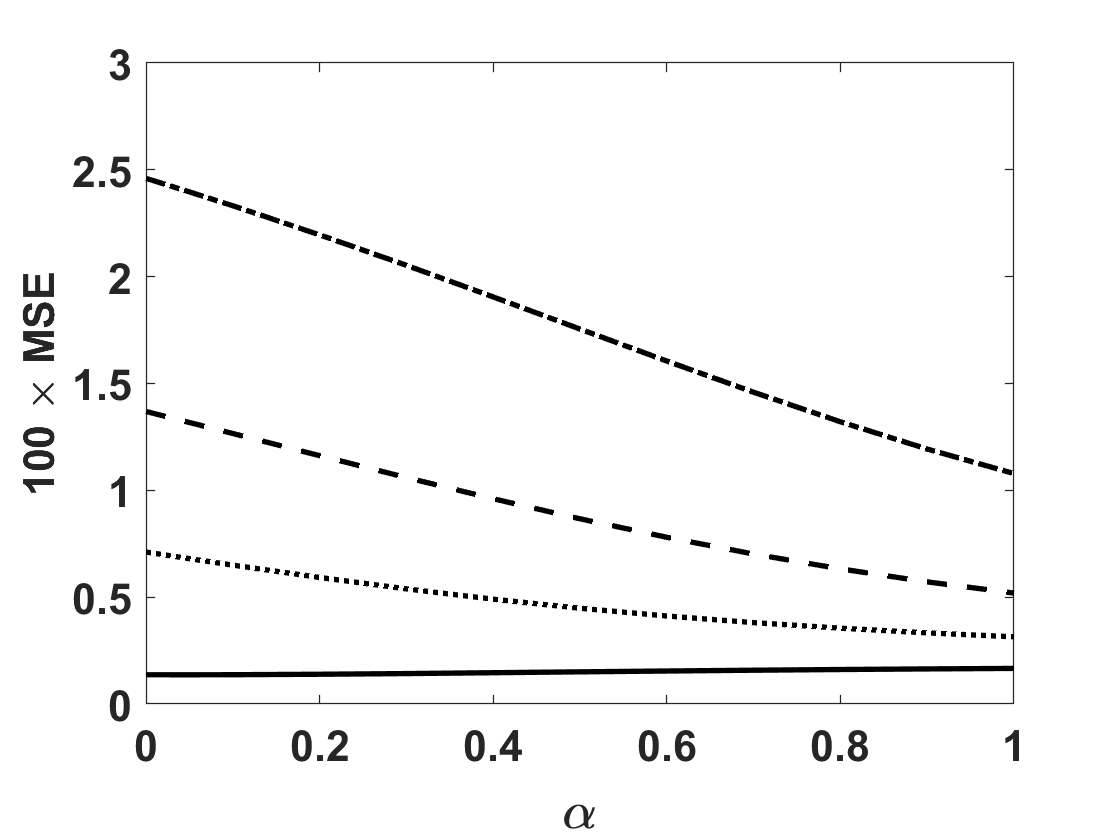}
		\label{FIG:Ex2_MSE_n100}}
	\caption{Empirical MSEs ($\times 100$) of the MDPDEs obtained from the simulation study conducted for Example 1 
		with $K=10$, $\theta=0.25$ and different contamination proportions
		[solid line: 0\%, dotted line: 10\%, dashed line: 15\%, dash-dotted line: 20\%]}
	\label{FIG:Ex2_MSE}
\end{figure}

In our simulation study, sample observations are generated from the Markov chain model
having transition probability matrix as in (\ref{EX2:P}) with $K=10$ and $\theta=0.25$. 
For different values of $T$, we simulate observed path of length $(T+1)$ with $X_0=1$
and compute the MDPDEs of $\theta$ for different $\alpha\geq 0$. 
We replicate this experiment 1000 times to compute the empirical mean squared error (MSE) of 
the MDPDEs with respect to its true value (0.25) for each $\alpha$. 
Further, to examine the robustness, a certain percentage, say $100\epsilon\%$, of the sample path is 
randomly replaced by observations from another finite Markov chain which always move forward with probability one 
(i.e, $\theta=1$ in the present model transition matrix) and repeat the same experiment to compute the MSEs under data contamination. 
The resulting values of MSEs of the MDPDEs are presented in Figure \ref{FIG:Ex2_MSE} for $T=50, 100$
and for 10\%, 15\%, 20\% contamination proportions, along with the pure data results (0\% contamination).
Recall that the MDPDE at $\alpha=0$ is the MLE which provides the least MSE under pure data.
The MSE under pure data increases, but very slowly, as $\alpha$ increases, in consistence with their asymptotic AREs (Table \ref{TAB:Ex2_ARE}).
However, under contamination, the MSE of the MLE (at $\alpha=0$) is significantly higher and 
decreases sharply as $\alpha>0$ increases; the MSEs remain more stable at larger values of $\alpha$.
This clearly indicate the claimed robustness of our proposed MDPDEs at $\alpha>0$ 
and that the extent of robustness further increases with increasing values of $\alpha$.

\subsection{Example 2: A Multi-parameter Extension of Example 1}

We now further extend the Markov chain model described in Example 1, 
so that the probability defining parameter $\theta$ depends on the current state, 
leading to the transition matrix 
\begin{eqnarray}
\boldsymbol{P}(\boldsymbol{\theta}) = \begin{bmatrix}
\begin{array}{cccccccc}
0 		 & 1 		& 0 		& 0 	 & \cdots & 0  & 0 & 0 \\
(1-\theta_2)^2 & 2\theta_2(1-\theta_2) 		& \theta_2^2 	& 0 	 & \cdots & 0  & 0 & 0\\
0 		 & (1-\theta_3)^2 & 2\theta_3(1-\theta_3)	& \theta_3^2 & \cdots & 0  & 0 & 0 \\
: 		 & 		: 	& 	: 		&  :	 &  \ddots & :  & : \\
0 		 & 	0		& 	0		& 0		& \cdots & (1-\theta_{K-1})^2 & 2\theta_{K-1}(1-\theta_{K-1})	& \theta_{K-1}^2  \\
0 		 &    0		&  	0		&  0     & \cdots  & 0 & 1  & 0 \\
\end{array}
\end{bmatrix},
\label{EX3:P}
\end{eqnarray}
with each $\theta_i\in [0,1]$ for $i=2, \ldots, K-1$.
Important models for explaining diffusion between gases or liquids are special cases of this Markov chain; 
see Section \ref{SEC:Ex_test} for an example.
Note that, here  the target parameter $\boldsymbol{\theta}=(\theta_2, \ldots, \theta_{K-1})^T$ is of dimension $d=(K-2)$,
but the individual components can be seen to be independent. 
The Markov chain corresponding to the transition matrix (\ref{EX3:P}) is also stationary and ergodic.
Its stationary distribution can be computed easily and is given by (with $\theta_1=1$, $\theta_K=0$ )
\begin{eqnarray}
\pi_{1o}&=& p_{1o}(\boldsymbol{\theta}) = \frac{1}{1+ \sum_{i=2}^K \theta_1^2\cdots\theta_{i-1}^2 (1-\theta_2)^{-2}(1-\theta_3)^{-2}\cdots(1-\theta_i)^{-2}},
\nonumber\\
\pi_{io}&=& p_{io}(\boldsymbol{\theta}) = \frac{\theta_1^2\cdots\theta_{i-1}^2 (1-\theta_2)^{-2}(1-\theta_3)^{-2}\cdots(1-\theta_i)^{-2}}{
1+ \sum_{2=1}^K \theta_1^2\cdots\theta_{i-1}^2 (1-\theta_2)^{-2}(1-\theta_3)^{-2}\cdots(1-\theta_i)^{-2}}, ~~~~i=2,3, \ldots, K.
\nonumber
\end{eqnarray}

Firstly, noting the similarity with Example 1, we can see that Assumptions (A1)--(A3) continue to hold for the transition matrix given by (\ref{EX3:P}) 
with $ C=\{ (1,2); (i, i+1), (i, i), (i, i-1) \mbox{ for } i=2, 3, \ldots, K-1; (K, K-1) \}$.
Then, given an observed  sequence $\mathcal{X}_T = \{X_0, X_1, \ldots, X_T\}$, one can easily verify that the MDPDE 
$\widehat{\theta}_{i,\alpha}$  of $\theta_i$, for each $i=2, \ldots, K-1$,
can be obtained separately by solving the respective estimating equation given by
\begin{eqnarray}
&& \left[ \theta_i^{2\alpha-1} \nu_{i(i+1)} + 2^{\alpha-1}\theta_i^{\alpha-1}(1-\theta_i)^{\alpha-1}(1-2\theta_i) \nu_{ii}
-  (1-\theta_i)^{2\alpha-1} \nu_{i(i-1)}\right]
\nonumber\\
&& ~~~~~~~~~~=  \nu_{i+} \left[\theta^{2\alpha+1} + 2^\alpha\theta^\alpha(1-\theta)^\alpha(1-2\theta)
- (1-\theta)^{2\alpha+1}\right]. 
\label{EQ:Ex3_MDPDE_EstEq}
\end{eqnarray}
Further, applying Theorem \ref{THM:MDPDE_conv}, one can verify that the asymptotic distribution of the $(K-2)$ dimensional 
MDPDE $\widehat{\boldsymbol{\theta}}_\alpha= (\widehat{\theta}_{i,\alpha} : i=2, \ldots, K-1)^t$ 
at the model with true parameter value $\boldsymbol{\theta}_0=({\theta}_{i0} : i=2, \ldots, K-1)^T$ is given by 
\begin{eqnarray}
\sqrt{T}\left(\widehat{\boldsymbol{\theta}}_{\alpha} - \boldsymbol{\theta}_0\right)
\mathop{\rightarrow}^\mathcal{D} 
\mathcal{N}_{K-2}\left(\boldsymbol{0}_d, \mbox{Diag}\left\{\sigma_\alpha^{(i)}(\boldsymbol{\theta}_0) : i=2, \ldots, K-1\right\}\right),
~~~~\mbox{as } T\rightarrow\infty, 
\nonumber
\end{eqnarray}
with 
$
\sigma_\alpha^{(i)}(\boldsymbol{\theta})=[1-p_{1o}(\boldsymbol{\theta}) - p_{Ko}(\boldsymbol{\theta})]^{-1}
\frac{V_{2,\alpha}(\theta_i)}{4 V_{1,\alpha}(\theta_i)^2},
$
where $V_{1,\alpha}$ and $V_{2,\alpha}$ are as defined in (\ref{EQ:V1}) and (\ref{EQ:V2}), respectively, 
but $p_{io}(\boldsymbol{\theta})$ is specific to the present example (as computed above).
Note that, the AREs of the MDPDEs of each parameter component in the present case are exactly the same as 
studied in Example 1 (Table \ref{TAB:Ex2_ARE}).
Their finite sample robustness advantages are also observed to have a similar pattern as in Example 1 (Figure \ref{FIG:Ex2_MSE}) 
via simulations and, hence, they are not reported here for brevity.

\subsection{Example 3: A Markov Chain for Epidemic Modeling}

Our final example would be another practically important Markov chain model,
namely the Greenwood model, for epidemic modeling of a contagious disease in a population of fixed size (say $K$, often small).
Suppose that every individual in the population can be in either of two categories, namely infected or uninfected,
and the disease evolves in some discrete time unit (a constant latent time period for a person to get infected).
Let $X_t$ denote the number of individuals who are still uninfected at time point $t =0, 1, 2, \ldots, T$,
so that $X_0=K$. If $\theta\in(0,1)$ denote the probability of contact between two individuals (one uninfected and one infected)
to produce a new infection at any time point, then $X_t$ is a Markov chain \citep{Gani/Jerwood:1971,Iosifescu:2007}    
with finite state-space $\mathcal{S}=\{0, 1, 2, \ldots, K\}$ and the transition matrix 
$$
\boldsymbol{P}(\theta) = \begin{bmatrix}
\begin{array}{cccccc}
1 		 & 0 		& 0 		 	 & \cdots & 0  & 0 \\
\theta & (1-\theta) 		& 0 	 	 & \cdots & 0  & 0 \\
\theta^2  & 2\theta(1-\theta) & (1-\theta_3)^2	 & \cdots & 0  & 0  \\
: 		 & 		: 	& 	: 		&    \ddots & :  & : \\
\theta^K &   \binom{K}{1}\theta^{K-1}(1-\theta)		&  	\binom{K}{1}\theta^{K-2}(1-\theta)^2    & \cdots  & \binom{K}{K-1}\theta(1-\theta)^{K-1} 
& (1-\theta)^K \\
\end{array}
\end{bmatrix}.
$$
Note that the underlying Markov chain can easily be seen to be stationary and ergodic.
Also Assumptions (A1)--(A3) hold with $ C=\{ (0,0); (i, 0), (i, 1), \ldots, (i, i) \mbox{ for } i=1, 2, \ldots, K \}$
so that $c=|C|=(K+1)(K+2)/2$.

Based on an observed  sequence $\mathcal{X}_T = \{X_0, X_1, \ldots, X_T\}$ from this model, 
we wish to estimate the target parameter $\theta$ (scalar).
We can use the MDPDE as a robust estimator of $\theta$
which can be obtained by solving the estimating equation (\ref{EQ:MDPDEF_EstEq}).
For the present case, the MDPDE estimating equation (\ref{EQ:MDPDEF_EstEq}) can be simplified as 
\begin{eqnarray}
&& \sum_{i=1}^{K}  \sum_{j=0}^i\binom{i}{j}^{\alpha}[(i-j)-i\theta]\theta^{(i-j)\alpha-1}(1-\theta)^{j\alpha-1} \nu_{ij}
\nonumber\\
&& ~~~~~~~~~~= \sum_{i=1}^{K} \nu_{i+}\sum_{j=0}^i\binom{i}{j}^{1+\alpha}[(i-j)-i\theta]\theta^{(i-j)(1+\alpha)-1}(1-\theta)^{j(1+\alpha)-1}. 
\label{EQ:Ex4_MDPDE_EstEq}
\end{eqnarray}
The asymptotic distribution of the resulting MDPDE, obtained by solving (\ref{EQ:Ex4_MDPDE_EstEq}), 
can again be calculated from Theorem \ref{THM:MDPDE_conv}, as in the previous examples;
we leave it for the readers as an exercise.

\begin{figure}[h]
	\centering
	\subfloat[$T=50$]{
		\includegraphics[width=0.45\textwidth]{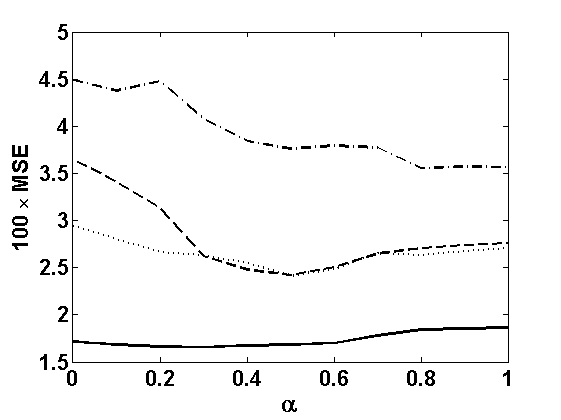}
		\label{FIG:Ex4_MSE_n50}}
	\subfloat[$T=100$]{
		\includegraphics[width=0.45\textwidth]{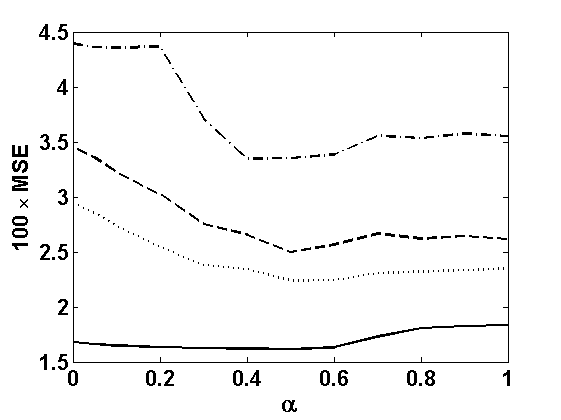}
		\label{FIG:Ex4_MSE_n100}}
	\caption{Empirical MSEs ($\times 100$) of the MDPDEs obtained from the simulation study conducted for Example 3 
		with $K=9$, $\theta=0.25$ and different contamination proportions
		[solid line: 0\%, dotted line: 10\%, dashed line: 15\%, dash-dotted line: 20\%]}
	\label{FIG:Ex4_MSE}
\end{figure}

We illustrate the finite sample performance of these MDPDEs and their claimed robustness by another simulation exercise.
We simulate observations of a fixed length ($T+1$) from the present Markov model with $K=9$ and $\theta=0.25$.
The empirical MSEs of the MDPDEs, obtained based on 1000 replications, are plotted in Figure \ref{FIG:Ex4_MSE} over $\alpha\in[0,1]$
for different amount of contaminations in the sample data. The contaminations are  incorporated in the sample path
at randomly selected location (say $i$) by taking the next step following Bin($i, 1$) distribution, 
i.e., by deterministically putting the next location to be $0$.
Note that, this type of contamination ends the chain there since this Markov chain can not go out of location 0 once it reaches there;
hence such a restriction may be considered as heavy contamination in the sample data.
Even under such a heavy contamination, we can see from Figure \ref{FIG:Ex4_MSE} that the proposed MDPDE with moderately large $\alpha>0$ 
provides significantly improved estimator (lower MSE) compared to the usual MLE (at $\alpha=0$).
Under pure data (no contamination), the MSEs show
a very mild increasing trend with increasing values of $\alpha$.
The MDPDEs with $\alpha$ around the value 0.5 provides the best trade-off in all the cases considered.

\section{Application of the MDPDE in Statistical Hypothesis Testing}
\label{SEC:Hyp_test}

\subsection{Wald-type Tests for General Composite Hypotheses}
\label{SEC:Hyp_test1}

Let us now consider the problem of testing statistical hypotheses about the underlying Markov chain 
defined in terms of the assumed parametric model $\mathcal{F}$. 
Under the set-up and notation of previous two sections, 
let us consider the general composite hypothesis given by 
\begin{eqnarray}
H_0: \boldsymbol{\theta}\in\Theta_0 ~~~~\mbox{ against }~~H_1: \boldsymbol{\theta}\notin\Theta_0, 
\label{EQ:Hyp_C}
\end{eqnarray}
where $\Theta_0$ is a pre-specified proper subset of  $\Theta$ having rank $r$.
In most applications, the null hypothesis in (\ref{EQ:Hyp_C}) can be re-expressed in terms of 
$r$ linearly independent restrictions of the form 
$$
\boldsymbol{h}(\boldsymbol{\theta}) =\boldsymbol{0}_r.
$$
Let us assume that the $d\times r$ matrix $\boldsymbol{H}\left(  \boldsymbol{{\boldsymbol\theta}}\right)  
=\frac{\partial\boldsymbol{h}(\boldsymbol{{\boldsymbol\theta}})}{\partial \boldsymbol{{\boldsymbol\theta}}}$
exists, has rank $r$ and is continuous in $\boldsymbol{{\boldsymbol\theta}}$. 
Further, if the assumptions of Theorem \ref{THM:MDPDE_conv} hold, 
there exists the MDPDE $\widehat{\boldsymbol{\theta}}_\alpha$
with tuning parameter $\alpha$ which satisfies the asymptotic normality result in (\ref{EQ:MDPDE_AN}).
We can use this MDPDE to construct a Wald-type test statistic for testing (\ref{EQ:Hyp_C})
as
\begin{equation}
W_{T,\alpha} = T \boldsymbol{h}^{t}\left(\widehat{\boldsymbol{\theta}}_\alpha\right)  
\left[  \boldsymbol{H}^{t}(\widehat{\boldsymbol{\theta}}_\alpha)
\boldsymbol{\Sigma}_\alpha(\boldsymbol{P}(\widehat{\boldsymbol{\theta}}_\alpha), \widehat{\boldsymbol{\theta}}_\alpha)
\boldsymbol{H}(\widehat{\boldsymbol{\theta}}_\alpha)\right]^{-1}
\boldsymbol{h}\left(\widehat{\boldsymbol{\theta}}_\alpha\right),
\label{EQ:Wald_TS_comp}
\end{equation}
where ${\boldsymbol{\boldsymbol\Sigma}}_\alpha$ is as defined in Theorem \ref{THM:MDPDE_conv}.
The asymptotic distribution of this proposed test statistic can be obtained from (\ref{EQ:MDPDE_AN})  which is presented in the following theorem.

\begin{theorem}\label{THM:Wald_Asym}
Assume that the conditions of Theorem \ref{THM:MDPDE_conv} hold true and the covariance matrix 
$\boldsymbol{\Sigma}_\alpha(\boldsymbol{P}(\boldsymbol{\theta}), \boldsymbol{\theta})$ is continuous 
in $\boldsymbol{\theta}$ around the null parameter values.
Then, under the null hypothesis in (\ref{EQ:Hyp_C}), 
the proposed Wald-type test statistic $W_{T,\alpha}$ asymptotically follows a chi-square distribution ($\chi_r^2$)
with $r$ degrees of freedom.
\end{theorem}

The above theorem can be used to obtain the asymptotic critical values for testing (\ref{EQ:Hyp_C}) 
based on $W_{T,\alpha}$ for all $\alpha\geq 0$.
Further properties of these proposed Wald-type tests can be easily obtained in the line of \cite{Ghosh/etc:2016} and \cite{Basu/etc:2018}.
In particular, the tests based on $W_{T,\alpha}$ are consistent, for all $\alpha\geq 0$, in the sense that their power
at any fixed alternative converges to one as $T\rightarrow\infty$.
It can also be shown that, under a contiguous sequence of alternative hypotheses 
$H_{1,T}:\boldsymbol{\theta} = \boldsymbol{{\boldsymbol\theta}}_{T}$, 
where $\boldsymbol{{\boldsymbol\theta}}_{T}=\boldsymbol{{\boldsymbol\theta}}_{0}+\frac{\boldsymbol{d}}{\sqrt{T}}$
with $\boldsymbol{d}\in \mathbb{R}^{d} \setminus \{\mathbf{0}_d\}$ and $\boldsymbol{\theta}_0\in\Theta_0$,
the Wald-type test statistic $W_{T,\alpha}$ asymptotically (as $T\rightarrow\infty$) follows a non-central chi-square distribution 
with degrees of freedom $r$  and non-centrality parameter 
$\delta_\alpha=\boldsymbol{d}^{t}\boldsymbol{H}\left(\boldsymbol{{\boldsymbol\theta}}_{0}\right)
\boldsymbol{\Sigma}_\alpha^\ast(\boldsymbol{\theta}_{0})^{-1}\boldsymbol{H}\left(\boldsymbol{{\boldsymbol\theta}}_{0}\right)^{t}\boldsymbol{d}$,
where $\boldsymbol{\Sigma}_\alpha^\ast(\boldsymbol{\theta})=\boldsymbol{H}^{t}(\boldsymbol{{\boldsymbol\theta}})
\boldsymbol{\boldsymbol\Sigma}_\alpha(\boldsymbol{P}(\boldsymbol{\theta}),\boldsymbol{\theta})\boldsymbol{H}(\boldsymbol{{\boldsymbol\theta}})$.
This result can be used to obtain the asymptotic power of the proposed Wald-type tests based on $W_{T,\alpha}$ under contiguous alternatives
and, hence,  to study their efficiency compared to any other consistent test.

\subsection{Robustness Analyses for the Wald-type tests}\label{SEC:IF_test}

The robustness of the proposed Wald-type tests based on $W_{T,\alpha}$ can also be theoretically justified 
through the concept of influence function analyses.
With the notation of Section \ref{SEC:IF_MDPDE}, let us first define the statistical functional 
corresponding to $W_{T,\alpha}$ at the true transition matrix $\boldsymbol{\Pi}$ as given by 
\begin{equation}
W_{\alpha}(\boldsymbol{\Pi})=\boldsymbol{h}^{t}(\boldsymbol{F}_\alpha(\boldsymbol{\Pi}))
\boldsymbol{\Sigma}_\alpha^\ast(\boldsymbol{F}_\alpha(\boldsymbol{\Pi}))^{-1}\boldsymbol{h}(\boldsymbol{F}_\alpha(\boldsymbol{\Pi})),
\label{EQ:WF}%
\end{equation}
where $\boldsymbol{F}_\alpha(\boldsymbol{\Pi})$ is the MDPDE functional with tuning parameter $\alpha$.
Then, we can define its first order influence function as 
$IF(\boldsymbol{t}; W_\alpha, \boldsymbol{\Pi})  = \left.\frac{\partial}{\partial\epsilon} W_\alpha(\boldsymbol{\Pi}_\epsilon)\right|_{\epsilon=0}$.
%
Since the influence function of a test statistics is examined at the null hypothesis, 
let us consider a parameter value $\boldsymbol{\theta}_0\in\Theta_0$ so that 
$\boldsymbol{h}(\boldsymbol{\theta}_0) = \boldsymbol{0}$ and $\boldsymbol{F}_\alpha(\boldsymbol{\Pi}^o)=\boldsymbol{\theta}_0$ with
$\boldsymbol{\Pi}^o=\boldsymbol{P}(\boldsymbol{\theta}_0)$.
Then, straightforward differentiation yields the first order IF as 
\begin{eqnarray}
IF(\boldsymbol{t}; W_\alpha, \boldsymbol{\Pi}^o)  &=& 2\boldsymbol{h}^{t}(\boldsymbol{F}_\alpha(\boldsymbol{\Pi}^o))
\boldsymbol{\Sigma}_\alpha^\ast(\boldsymbol{F}_\alpha(\boldsymbol{\Pi}^o))^{-1}\boldsymbol{H}^t(\boldsymbol{F}_\alpha(\boldsymbol{\Pi}^o))
IF(\boldsymbol{t}; \boldsymbol{F}_\alpha, \boldsymbol{\Pi}^o)  
\nonumber\\
&&~~+  \boldsymbol{h}^{t}(\boldsymbol{F}_\alpha(\boldsymbol{\Pi}^o))
\left[\frac{\partial}{\partial\epsilon}\boldsymbol{\Sigma}_\alpha^\ast(\boldsymbol{F}_\alpha(\boldsymbol{\Pi}_\epsilon))^{-1}\right]_{\epsilon=0}
\boldsymbol{h}(\boldsymbol{F}_\alpha(\boldsymbol{\Pi}^o)),
\nonumber\\
&=&0.
\end{eqnarray}

Since this first order IF is always zero, it is non-informative to indicate the robustness of the test procedure and,  
hence, we need to consider the second order IF of the test functional $W_\alpha$.
By another round of differentiation, we get 
\begin{eqnarray}
IF_2(\boldsymbol{t}; W_\alpha, \boldsymbol{\Pi}^o) 
&=& \left.\frac{\partial^2}{\partial\epsilon^2} W_\alpha(\boldsymbol{\Pi}_\epsilon)\right|_{\epsilon=0} 
\nonumber\\&=& 2 IF(\boldsymbol{t}; \boldsymbol{F}_\alpha, \boldsymbol{\Pi}^o)^t\boldsymbol{H}(\boldsymbol{\theta}_0)
\boldsymbol{\Sigma}_\alpha^\ast(\boldsymbol{\theta}_0)^{-1}\boldsymbol{H}^{t}(\boldsymbol{\theta}_0)
IF(\boldsymbol{t}; \boldsymbol{F}_\alpha, \boldsymbol{\Pi}^o).
\end{eqnarray}
Note that this second order IF of the Wald-type test functional $W_\alpha$ depends directly on the IF of 
the underlying MDPDE, used in the construction of the test statistics.
Therefore, the robustness properties of the MDPDE based Wald-type tests are exactly similar to those of the MDPDE
as discussed in Section \ref{SEC:IF_MDPDE}.

We can also study the IF for the asymptotic level and power of the proposed Wald-type tests
in the line  of \cite{Ghosh/etc:2016} and \cite{Basu/etc:2018}, which would be a linear function of the IF of the underlying MDPDE.
Hence, their robustness would also be implied by the robustness of the MDPDE used in the test statistics 
(i.e., greater robustness with increasing $\alpha>0$).

\subsection{An Example: Test for the Bernoulli-Laplace Model of Diffusion}
\label{SEC:Ex_test}

The Markov chain associated with the famous Bernoulli-Laplace diffusion model 
for two incompressible gases or liquids between two containers \citep{Iosifescu:2007}
is defined by the $K\times K$ transition matrix 
$$
\boldsymbol{P}_{\ast} = \begin{bmatrix}
\begin{array}{cccccccc}
r_1 		 & p_1 		& 0 		& 0 	 & \cdots & 0  & 0 & 0 \\
q_2 & r_2 		& p_2 	& 0 	 & \cdots & 0  & 0 & 0\\
0 		 & q_3 & r_3	& p_3 & \cdots & 0  & 0 & 0 \\
: 		 & 		: 	& 	: 		&  :	 &  \ddots & :  & : \\
0 		 & 	0		& 	0		& 0		& \cdots & q_{K-1} & r_{K-1}	& p_{K-1}  \\
0 		 &    0		&  	0		&  0     & \cdots  & 0 & q_K  & r_K \\
\end{array}
\end{bmatrix},
$$
where $K>1$ and 
$$
p_i = \left(\frac{K-i}{K-1}\right)^2, 
~~ q_i = \left(\frac{i-1}{K-1}\right)^2,
~~ r_i = 2\left(\frac{K-i}{K-1}\right)\left(\frac{i-1}{K-1}\right),
~~~ i =1, 2, \ldots, K.  
$$
Suppose that,  given a sequence $\mathcal{X}_T = \{X_0, X_1, \ldots, X_T\}$  observed from a suitable process,
we wish to test if it satisfies the Bernoulli-Laplace model, i.e., if it is generated according to the above transition matrix. 
As a convenient class of alternatives, we may consider the family $\mathcal{F}$ of parametric transition matrices 
of the form (\ref{EX3:P}) which was discussed in Section \ref{SEC:Ex}.2. 
Note that the transition matrix $\boldsymbol{P}_{\ast}$ of the Bernoulli-Laplace model belongs to this parametric family $\mathcal{F}$ 
for the parameter value $\theta_i = (K-i)/(K-1)$ for each $i=2, \ldots, K-1$. 
Then, our targeted hypothesis can be expressed as a (simple) parametric hypothesis (within $\mathcal{F}$) given by
\begin{eqnarray}
H_0 : \theta_i = \frac{(K-i)}{(K-1)}, ~i=2, \ldots, K-1,
~~\mbox{ against }~~ H_1: H_0 \mbox{ is not true}.
\label{EQ:Ex3_Hyp}
\end{eqnarray}

Clearly this hypothesis (\ref{EQ:Ex3_Hyp}) belongs to the class of hypotheses considered in Section \ref{SEC:Hyp_test1}
with $r=K-2$,  $\boldsymbol{h}(\boldsymbol{\theta}) = \left(\theta_i - \frac{(K-i)}{(K-1)} : ~i=2, \ldots, K-1\right)$
and hence $\boldsymbol{H}(\boldsymbol{\theta}) =  \boldsymbol{I}_{K-2}$, the identity matrix of dimension $(K-2)$.
So, using the (asymptotic) properties of the MDPDEs of $\boldsymbol{\theta}$ derived in Example 2 (Section \ref{SEC:Ex}.2)
and the theory discussed Section \ref{SEC:Hyp_test1}, 
we can construct a robust Wald-type test statistic as defined in (\ref{EQ:Wald_TS_comp}) for testing the hypothesis is (\ref{EQ:Ex3_Hyp}). 
In this case, through the insertion of the particular forms of $\boldsymbol{h}$ and $\boldsymbol{H}$,
our proposed test statistic is simplified to be 
\begin{equation}
W_{T,\alpha} = T \sum_{i=2}^{K-1}\frac{\left(\widehat{{\theta}}_\alpha - \frac{K-i}{K-1}\right)^2}{\Sigma_\alpha(\boldsymbol{P}(\theta_{i0}), {\theta}_{i0})}.
\label{EQ:Ex3_Wald_TS}
\end{equation}
It is easy to see that $W_{T,\alpha}$ coincides with the usual Wald test at $\alpha=0$, and provide a robust generalization at $\alpha>0$.
Further, for any $\alpha\geq 0$, by Theorem \ref{THM:Wald_Asym} and the continuity of the asymptotic variance matrix 
$\Sigma_\alpha(\boldsymbol{P}(\theta), {\theta})$ of the MDPDEs of $\boldsymbol{\theta}$,
the test statistic in (\ref{EQ:Ex3_Wald_TS}) asymptotically has a $\chi_{K-2}^2$ distribution as $T\rightarrow\infty$,
under the null hypothesis in (\ref{EQ:Ex3_Hyp}). 
Therefore, we reject the null hypothesis in (\ref{EQ:Ex3_Hyp}) at $\zeta$ level of significance if 
$$
W_{T,\alpha} > \chi^2_{K-2, \zeta}, \mbox{ the $(1-\zeta)$-th quantile of $\chi_{K-2}^2$ distribution}.
$$


\subsection{Test for the Similarity of Two Sequences of Markov chains}
\label{SEC:two_sample}

We now discuss the problem of comparing two Markov chain sequences.
%
Suppose that we observe two independent sequences $\mathcal{X}^{(j)} = \{X_0^{(j)}, X_1^{(j)}, \ldots, X_{T_j}^{(j)}\}$ of length $(T_j+1)$, 
for $j=1,2$, from Markov chains  having the same finite state-space $\mathcal{S}=\{1, 2, \ldots, K\}$ and 
transition probabilities  belonging to the parameter family 
$\mathcal{F} = \left\{\boldsymbol{P}(\boldsymbol{\theta}) = (p_{ij}(\boldsymbol{\theta}))_{i,j=1, \ldots,K}
: \boldsymbol{\theta}\in\Theta\subseteq\mathbb{R}^d \right\}$. 
Our aim is to statistically test if the two sequences are generated from the same Markov chain. 
If we assume that the transition matrix of $\mathcal{X}^{(j)}$ is $\boldsymbol{P}(\boldsymbol{\theta}_j)$
for some $\boldsymbol{\theta}_j\in\Theta$, $j=1, 2$, then our problem corresponds to testing the hypothesis
\begin{eqnarray}
H_0: \boldsymbol{\theta}_1 = \boldsymbol{\theta}_2 ~~~~\mbox{ against }~~H_1: \boldsymbol{\theta}_1 \neq \boldsymbol{\theta}_2. 
\label{EQ:Hyp_2s}
\end{eqnarray}

In order to develop a robust test statistic for testing (\ref{EQ:Hyp_2s}),
let us denote by  $\widehat{\boldsymbol{\theta}}_\alpha^{(j)}$ the MDPDE of $\boldsymbol{\theta}_j$, with tuning parameter $\alpha$,
obtained based on the sequence $\mathcal{X}^{(j)}$,  for $j=1,2$, respectively. 
These two MDPDEs are then independent and each of them is asymptotically normal, under the assumptions of Theorem \ref{THM:MDPDE_conv},
as $T_1, T_2 \rightarrow\infty$.
Accordingly, for testing (\ref{EQ:Hyp_2s}), we may consider a Wald-type test statistic defined based on these MDPDEs  as
\begin{equation}
W_{T_1, T_2}^{(\alpha)} = T_1 T_2 \left(\widehat{\boldsymbol{\theta}}_\alpha^{(1)} - \widehat{\boldsymbol{\theta}}_\alpha^{(2)}\right)^t  
\left[  T_2\boldsymbol{\Sigma}_\alpha\left(\boldsymbol{P}(\widehat{\boldsymbol{\theta}}_\alpha^{(1)}), \widehat{\boldsymbol{\theta}}_\alpha^{(1)}\right)
+ T_1 \boldsymbol{\Sigma}_\alpha\left(\boldsymbol{P}(\widehat{\boldsymbol{\theta}}_\alpha^{(2)}), \widehat{\boldsymbol{\theta}}_\alpha^{(2)}\right)\right]^{-1}
\left(\widehat{\boldsymbol{\theta}}_\alpha^{(1)} - \widehat{\boldsymbol{\theta}}_\alpha^{(2)}\right).
\nonumber
\label{EQ:Wald_TS_2s}
\end{equation}
The critical values for testing (\ref{EQ:Hyp_2s}) using the test statistic $W_{T_1, T_2}^{(\alpha)}$ 
can be obtained from its asymptotic distribution, which is presented in the following theorem.

\begin{theorem}\label{THM:Wald_Asym2s}
Assume that the conditions of Theorem \ref{THM:MDPDE_conv} hold true and the covariance matrix 
$\boldsymbol{\Sigma}_\alpha(\boldsymbol{P}(\boldsymbol{\theta}), \boldsymbol{\theta})$ is continuous 
in $\boldsymbol{\theta}$ around the null parameter value $\boldsymbol{\theta}_1=\boldsymbol{\theta}_2$.
Suppose that $T_1, T_2 \rightarrow\infty$ in such a way that $T_1/(T_1+T_2) \rightarrow w$ for some $w\in(0,1)$.
Then, under the null hypothesis in (\ref{EQ:Hyp_2s}), the asymptotic distribution of
our MDPDE based Wald-type test statistic $W_{T_1,T_2}^{(\alpha)}$ is $\chi_d^2$.
\end{theorem}

We can derive other properties of the test based on $W_{T_1,T_2}^{(\alpha)}$ 
by extending the theory of usual two-sample Wald-type tests in the line of \cite{Ghosh/etc:2018}.
In particular, this  test is also consistent for all $\alpha\geq 0$ 
and its power under pure data as well as its robustness depend directly 
on the relative efficiency and the robustness of the MDPDE used in constructing the test statistics.

\section{Further Extensions}\label{SEC:Extn}

\subsection{Multiple Sequences of Markov Chain Observations}\label{SEC:Extn1}

Extending the notation from Section \ref{SEC:two_sample}, let us now consider $n (\geq 2)$ sequences of Markov chain observations 
denoted by $\mathcal{X}^{(j)} = \{X_0^{(j)}, X_1^{(j)}, \ldots, X_{T_j}^{(j)}\}$, for $j=1,2, \ldots, n$.
Suppose that all of them are generated from the same Markov chain having finite state-space $\mathcal{S}=\{1, 2, \ldots, K\}$
and transition probability matrix $\boldsymbol{\Pi}$ to be modeled by $\boldsymbol{P}(\boldsymbol{\theta})$.
Our aim is to estimate the parameter $\boldsymbol{\theta}$ from the combined information from all these $n$ sequences of observations.

Here, we define the (non-parametric) probability estimates $\widehat{\pi}_{ij}$ and $\widehat{\pi}_{io}$ 
from the combined (average) frequency counts $\nu_{ij}^{(n)}$ and $\nu_{i+}^{(n)}$ obtained from all the $n$ chains in place of $\nu_{ij}$ and $\nu_{i+}$,
respectively, in (\ref{EQ:Est_np}), where
\begin{eqnarray}
\nu_{ij}^{(n)}=\frac{1}{n}\sum_{j=1}^n \sum_{t=0}^{T_j-1}I(X_t^{(j)}=i,X_{t+1}^{(j)}=j),
~~~~
\nu_{i+}^{(n)}=\sum_{j=1}^K \nu_{ij}^{(n)},
~~~i,j=1, \ldots, K.
\label{EQ:freq_countn}
\end{eqnarray}
Then, we can proceed exactly as in the case of one sequence, described in Section \ref{SEC:MDPDE_Est},
with these new definitions of $\widehat{\pi}_{ij}$ and $\widehat{\pi}_{io}$ to define the MDPDE of $\boldsymbol{\theta}$
with tuning parameters $\alpha$. 
The robustness analyses of Section \ref{SEC:IF_MDPDE} would also be valid in this case.
However, the asymptotic results derived in Section \ref{SEC:MDPDE_asymp} are needed to be modified appropriately for the present case. 
For simplicity in discussion, in the following, we will assume that $T_1=T_2=\ldots=T_n=T$;
the results can be easily extended for the case of different $T_j$s.

Note that, there can be two directions of asymptotic derivation.
Firstly, when the number of sequences ($n$) is a small finite number and the length of the sequences $T\rightarrow\infty$,
the main asymptotic results (\ref{EQ:CLT}) can be modified easily leading to the result
\begin{eqnarray}
\boldsymbol{\eta} := \sqrt{nT}\left(\widehat{\boldsymbol{\Pi}}_{C} - \boldsymbol{\Pi}_C\right)
\mathop{\rightarrow}^\mathcal{D} \mathcal{N}_c\left(\boldsymbol{0}_c, \boldsymbol{\Lambda}(\boldsymbol{\Pi})\right),
~~~~\mbox{as } T\rightarrow\infty,
\label{EQ:CLT2}
\end{eqnarray}
where $\widehat{\boldsymbol{\Pi}}=(\widehat{\pi}_{ij})$ is now defined from the modified (average) frequency counts 
$\nu_{ij}^{(n)}$ and $\nu_{i+}^{(n)}$ given in (\ref{EQ:freq_countn}) and the asymptotic variance matrix  
$\boldsymbol{\Lambda}(\boldsymbol{\Pi})$ is exactly the same as defined in Section \ref{SEC:MDPDE_asymp}.

In the second practically relevant case, we may observe a large number ($n$) of sequences, each of which has a  small finite length $T$,
so that the asymptotics has to be done as $n\rightarrow\infty$.
This second type of asymptotics is studied in detail by \cite{Anderson/Goodman:1957} under finite Markov chains 
with two different initial conditions on $n_i = \sum_{j=1}^n I(X_0^{(j)}=i)$, the number 
of observations in state $i$ at time $t=0$, for each $i=1,2, \ldots, K$. 
For non-random $n_i$s one needs to assume that $n_i/n \rightarrow w_i\in(0,1)$ with $\sum_i w_i=1$, 
whereas for random $n_i$s they were assumed to have a multinomial distribution with probabilities $w_i$ and sample size $n$.  
In either cases, a modified version of  (\ref{EQ:CLT}) has been derived in \cite{Anderson/Goodman:1957} 
for a stationary ergodic chain starting from the stationary state,
which is exactly the same as  (\ref{EQ:CLT2}) but now as $n\rightarrow\infty$.

Therefore, in both the directions of asymptotics, we can use the same result (\ref{EQ:CLT2}) 
to derive the asymptotic distribution of our MDPDEs in the present case. 
Proceeding exactly as in the proof of Theorem \ref{THM:MDPDE_conv_Out}, we now have the following result
\begin{eqnarray}
\sqrt{nT}\left(\widehat{\boldsymbol{\theta}}_{\alpha} - \boldsymbol{\theta}_0\right)
\mathop{\rightarrow}^\mathcal{D} 
\mathcal{N}_d\left(\boldsymbol{0}_d, \boldsymbol{\Sigma}_\alpha(\boldsymbol{P}^o, \boldsymbol{\theta}_0)\right),
\label{EQ:MDPDE_AN2}
\end{eqnarray} 
under both directions of asymptotic, i.e., either as $T\rightarrow\infty$ or as $n\rightarrow\infty$.
Therefore, all the subsequent properties of the MDPDE and the associated Wald-type tests can also be extended 
in the present case of multiple sequence of Markov chain observations in a straightforward manner.

\subsection{Finite Markov Chains of Higher Order}

Our proposed statistical methodologies can also be extended for higher order Markov chains.
Let us now illustrate it for a second order Markov chain sequence $\mathcal{X}_T = \{X_0, X_1, \ldots, X_T\}$  
having finite state-space $\mathcal{S}=\{1, 2, \ldots, K\}$ and stationary transition probabilities
$$
\pi_{ijl} = P(X_t=l | X_{t-1}=j, X_{t-2}=i), ~~~~~~i,j,l=1, 2, \ldots, K.
$$ 
Suppose we wish to model these transition probabilities by some parametric family of $K\times K\times K$ transition (3D)-matrices 
$\widetilde{\mathcal{F}} = \left\{\widetilde{\boldsymbol{P}}(\boldsymbol{\theta}) = (p_{ijl}(\boldsymbol{\theta}))_{i,j,l=1, \ldots,K}
: \boldsymbol{\theta}\in\Theta\subseteq\mathbb{R}^d \right\}$ so that our objective is to estimate the unknown parameter $\boldsymbol{\theta}$
from the observed sequence.

To define the MDPDE of $\boldsymbol{\theta}$ for this second order Markov chain, 
we re-express it as a first order Markov chain with the state-space $\mathcal{S}\times\mathcal{S} = \{(i,j): i,j=1, 2, \ldots, K\}$.
This resulting first order Markov chain over $\mathcal{S}\times \mathcal{S}$ will have a $K^2\times K^2$
transition matrix of the form $\boldsymbol{\Pi}=(\pi_{(i,j)(h,l)})_{(i,j), (h,l)\in \mathcal{S}\times\mathcal{S}}$, 
where $\pi_{(i,j)(h,l)}= \delta_{jh}\pi_{ijl}$. The parametric model family $\widetilde{\mathcal{F}}$ can also 
be converted similarly to a parametric family of $K^2\times K^2$ transition matrices given by
$$
\mathcal{F}=\left\{ {\boldsymbol{P}}(\boldsymbol{\theta}) = ((p_{(i,j)(h,l)}(\boldsymbol{\theta})))_{(i,j), (h,l)\in \mathcal{S}\times\mathcal{S}}
: \boldsymbol{\theta}\in\Theta\subseteq\mathbb{R}^d \right\},
~~~~
\mbox{ where }
~p_{(i,j)(h,l)}(\boldsymbol{\theta})=\delta_{jh}p_{ijl}(\boldsymbol{\theta}).
$$
Then, the MDPDE of $\boldsymbol{\theta}$ can be defined as in \ref{SEC:MDPDE_Est} by minimizing the appropriate DPD measure between the 
modified model transition matrix ${\boldsymbol{P}}(\boldsymbol{\theta})$ and the (non-parametric) estimate of $\boldsymbol{\Pi}$.
In this case, the non-parametric estimates of the original transition probabilities $\pi_{ijl}$s are given by 
$$
\widehat{\pi}_{i,j,l} = \frac{\sum_{t=2}^{T}I(X_{t-2}=i,X_{t-1}=j,X_t=l)}{\sum_{l=1}^K\sum_{j=1}^K\sum_{t=2}^{T}I(X_{t-2}=i,X_{t-1}=j,X_t=l)},
~~~~~ i,j, l = 1, 2, \ldots, K.
$$
Therefore, an estimate of the element of the modified transition matrix $\boldsymbol{\Pi}$ is given by 
$$
\widehat{\pi}_{(i,j)(h,l)}=\delta_{jh}\widehat{\pi}_{ijl} =  \frac{\delta_{jh}\sum_{t=2}^{T}I(X_{t-2}=i,X_{t-1}=j,X_t=l)}{\sum_{l=1}^K\sum_{j=1}^K\sum_{t=2}^{T}I(X_{t-2}=i,X_{t-1}=j,X_t=l)},
~~~~~~(i,j), (h,l)\in \mathcal{S}\times\mathcal{S},
$$
and the estimate of the associated stationary probabilities $\pi_{(i,j)o}$ are given by 
$$
\widehat{\pi}_{(i,j)o} = \frac{1}{T}\sum_{l=1}^K\sum_{h=1}^K\widehat{\pi}_{(i,j)(h,l)}
= \frac{1}{T}\sum_{l=1}^K\widehat{\pi}_{ijl},
~~~~~ (i,j)\in \mathcal{S}\times\mathcal{S}.
$$
Then, the MDPDE of $\boldsymbol{\theta}$ is defined as a minimizer of the objective function in (\ref{EQ:MDPDE_objFunc}),
which now reads as 
\begin{eqnarray}
H_{T,\alpha}^{(2)}(\boldsymbol{\theta}) &=& \frac{1}{1+\alpha}\sum_{(i,j)\in \mathcal{S}\times\mathcal{S}}  \widehat{\pi}_{(i,j)o} 
\sum_{(h,l)\in \mathcal{S}\times\mathcal{S}} 
\left\{p_{(i,j)(h,l)}(\boldsymbol{\theta})^{1+\alpha} - \left(1 + \frac{1}{\alpha}\right)  
p_{(i,j)(h,l)}(\boldsymbol{\theta})^\alpha \widehat{\pi}_{(i,j)(h,l)}\right\}
\nonumber\\
&=& \frac{1}{1+\alpha}\sum_{i=1}^K\sum_{j=1}^K  \widehat{\pi}_{(i,j)o} 
\sum_{l=1}^K \left\{p_{ijl}(\boldsymbol{\theta})^{1+\alpha} - \left(1 + \frac{1}{\alpha}\right)  
p_{ijl}(\boldsymbol{\theta})^\alpha \widehat{\pi}_{ijl}\right\}.
\label{EQ:MDPDEF_objFuncH2}
\end{eqnarray}
In analogue to (\ref{EQ:MDPDE_EstEq}), the estimation equation of the MDPDE will now have the form
\begin{eqnarray}
\boldsymbol{U}_{T,\alpha}^{(2)}(\boldsymbol{\theta}) := \sum_{i=1}^K\sum_{j=1}^K  \widehat{\pi}_{(i,j)o}  \sum_{l=1}^K 
\boldsymbol{\psi}_{ijl}(\boldsymbol{\theta})
\left(p_{ijl}(\boldsymbol{\theta})- \widehat{\pi}_{ijl}\right)p_{ijl}(\boldsymbol{\theta})^{\alpha}  = \boldsymbol{0}_d.
\label{EQ:MDPDE_EstEqH2}
\end{eqnarray}
All the asymptotic properties of the resulting MDPDE  can be easily obtained from the results of Section \ref{SEC:MDPDE_asymp} 
via the first order Markov chain representation over $\mathcal{S}\times\mathcal{S}$.
Then, the subsequent testing procedures can also be developed in a similar fashion.

Note that, the objective function (\ref{EQ:MDPDEF_objFuncH2}) and the estimating equation (\ref{EQ:MDPDE_EstEqH2}),
corresponding to the MDPDE in a second order Markov chain, have quite general structures 
that can easily be extended for Markov chains of any higher order.
Suppose we have a sequence $\mathcal{X}_T = \{X_0, X_1, \ldots, X_T\}$  of observations from a Markov chain of order $r (\geq 2)$
having finite state-space $\mathcal{S}=\{1, 2, \ldots, K\}$ and stationary transition probabilities
$$
\pi_{i_1i_2\cdots i_{r+1}} = P(X_t=i_{r+1}|X_{t-r}=i_1, X_{t-r+1}=i_{2}, \cdots, X_{t-1}=i_{r}), ~~~~i_j=1,  \ldots, K; ~j=1,  \ldots, (r+1).
$$ 
If these transition probabilities are modeled by some parametric model of the form $p_{i_1i_2\cdots i_{r+1}}(\boldsymbol{\theta})$, 
the MDPDE of the corresponding parameter $\boldsymbol{\theta}$ would be defined as the minimizer of the objective function 
\begin{eqnarray}
H_{T,\alpha}^{(r)}(\boldsymbol{\theta}) = \frac{1}{1+\alpha}\sum_{i_1=1}^K\cdots\sum_{i_{r}=1}^K  \widehat{\pi}_{(i_1,\cdots, i_{r})o} 
\sum_{i_{r+1}=1}^K \left\{p_{i_1i_2\cdots i_{r+1}}(\boldsymbol{\theta})^{1+\alpha} - \left(1 + \frac{1}{\alpha}\right)  
p_{i_1i_2\cdots i_{r+1}}(\boldsymbol{\theta})^\alpha \widehat{\pi}_{i_1i_2\cdots i_{r+1}}\right\},
\nonumber
\end{eqnarray}
or the solution of the estimating equation
\begin{eqnarray}
\boldsymbol{U}_{T,\alpha}^{(r)}(\boldsymbol{\theta}) := \sum_{i_1=1}^K\cdots\sum_{i_{r}=1}^K  \widehat{\pi}_{(i_1,\cdots, i_{r})o}\sum_{i_{r+1}=1}^K 
\boldsymbol{\psi}_{i_1i_2\cdots i_{r+1}}(\boldsymbol{\theta})
\left(p_{i_1i_2\cdots i_{r+1}}(\boldsymbol{\theta})- \widehat{\pi}_{i_1i_2\cdots i_{r+1}}\right)p_{i_1i_2\cdots i_{r+1}}(\boldsymbol{\theta})^{\alpha}  
= \boldsymbol{0}_d,
\nonumber
\label{EQ:MDPDE_EstEqH}
\end{eqnarray}
where
\begin{eqnarray}
\widehat{\pi}_{i_1i_2\cdots i_{r+1}} &=& \frac{\sum_{t=r}^T I(X_{t-r}=i_1, X_{t-r+1}=i_{2}, \cdots, X_t=i_{r+1})}{
\sum_{i_1=1}^K\cdots\sum_{i_{r}=1}^K \sum_{t=r}^T I(X_{t-r}=i_1, X_{t-r+1}=i_{2}, \cdots, X_t=i_{r+1})}, 
\nonumber\\
\widehat{\pi}_{(i_1, \cdots, i_{r})o} &=& \frac{1}{T}\sum_{i_{r+1}=1}^K \widehat{\pi}_{i_1i_2\cdots i_{r+1}}, ~~~~\mbox{ for }
i_j=1, 2, \ldots, K; ~ j=1, 2, \ldots, r.
\nonumber
\end{eqnarray}
Further in-depth investigations of the MDPDEs under such higher order Markov chains as well as their applications
would be interesting future research.  
Among others, an important application would be the robust test for the order of a Markov chain based on the given observations,
i.e., to test if the given sequence of observations is coming from a Markov chain of order $r$ via the hypothesis
$H_0~:~\pi_{i_1i_2\cdots i_{r+1}} =\pi_{i_2\cdots i_{r+1}}$ with  $i_j\in \{1,  \ldots, K\}$, for each $j=1,  \ldots, (r+1)$.
There have been several approaches for determination of the order of a Markov chain 
including the information criteria like AIC and BIC \citep{Zhao/etc:2001}, 
divergence based test of the above $H_0$ \citep{Menendez/etc:2001} and mutual information based procedures \citep{Papapetrou/Kugiumtzis:2013}.
In the same spirit, we can also use the MDPDE and the associated DPD measure to develop appropriate robust test 
for the above $H_0$ corresponding to the order selection of Markov chain observations which we hope to study in detail in a sequel work.

\subsection{Finite Markov Chains with Time-dependent Transition probabilities}

In several practical applications, the Markov chain may not be stationary, i.e., the transition probabilities depend on time.
We can also extend the concept of MDPDE for robust parameter estimation for such non-stationary cases.
First note that, if we have one large sequence of observations from a non-stationary Markov chain, without loss of generality, 
the observations far away from the starting point can often be treated as stationary, under appropriate assumptions, for the inference purpose. 
This is because, in many common situations, such a chain mixes very fast (exponential or geometric mixing)
and, hence, behave like a stationary chain in long run.
So, here, we discuss the extension of the MDPDE only for the case of 
several small-length sequence of observations from a non-stationary Markov chain model.

Let us consider the set-up and notation of Section \ref{SEC:Extn1}, where we have observed $n$ sequence of observations each of length $(T+1)$ 
(for simplicity), with large enough $n$ ($\rightarrow\infty$) and relatively small $T$.
Suppose that the underlying Markov chain has transition probabilities $\pi_{ij}(t)$ for a given time-point $t=0, 1, \ldots, T$,
and we model it by a parametric family of transition matrices depending on time $t$, i.e., by the family
$\mathcal{F} = \left\{\boldsymbol{P}(t;\boldsymbol{\theta}) = (p_{ij}(t;\boldsymbol{\theta}))_{i,j=1, \ldots,K}
: \boldsymbol{\theta}\in\Theta\subseteq\mathbb{R}^d \right\}$, 
where $p_{ij}(t;\boldsymbol{\theta})$ are known functions depending on the unknown $d$-dimensional parameter vector 
$\boldsymbol{\theta}=(\theta_1, \ldots, \theta_d)'\in \Theta$ for each time point $t$. 
We want to estimate $\boldsymbol{\theta}$ based on the observed sequences $\mathcal{X}_T^{(l)}$ for $l=1, 2, \ldots, n$.

In this context, the non-parametric MLEs of the transition probabilities $\pi_{ij}(t)$ are given by  
\begin{eqnarray}
\widehat{\pi}_{ij}^{(n)}(t)=\frac{\sum_{l=1}^n I(X_{t-1}^{(l)}=i,X_{t}^{(l)}=j)}{\sum_{j=1}^K \sum_{l=1}^n I(X_{t-1}^{(l)}=i,X_{t}^{(l)}=j)},
~~~i,j=1, \ldots, K; t=0,1, \ldots, T.
\nonumber
\end{eqnarray}
If $\widehat{\boldsymbol{\Pi}}^{(n)}(t) = (\widehat{\pi}_{ij}^{(n)}(t))_{i,j=1, 2, \ldots, K}$ denote the corresponding estimate of 
$\boldsymbol{\Pi}(t)$ for each $t=0, 1, \ldots, T$, then the MDPDE of $\boldsymbol{\theta}$ can be defined as the minimizer of 
an appropriate generalization of the DPD measures between $\widehat{\boldsymbol{\Pi}}^{(n)}(t)$  and $\boldsymbol{P}(t;\boldsymbol{\theta})$.
The most intuitive choice (extending the idea from \cite{Ghosh/Basu:2013}) is to minimize the total discrepancy measure 
$
\sum\limits_{t=0}^T \sum\limits_{i=1}^K \widehat{\pi}_{io}^{(n)}(t) 
\cdot d_\alpha(\widehat{\boldsymbol{\Pi}}_i^{(n)}(t),\boldsymbol{P}_i(t;\boldsymbol{\theta})),
$
where $\widehat{\boldsymbol{\Pi}}_i^{(n)}(t)$ and $\boldsymbol{P}_i(t;\boldsymbol{\theta})$ denote the $i$-th row of 
$\widehat{\boldsymbol{\Pi}}^{(n)}(t)$  and $\boldsymbol{P}(t;\boldsymbol{\theta})$, respectively, and 
$$
\widehat{\pi}_{io}^{(n)}(t)= \frac{1}{n}\sum_{j=1}^K \sum_{l=1}^n I(X_{t-1}^{(l)}=i,X_{t}^{(l)}=j),
~~~i=1,2, \ldots, K; ~ t=0,1, \ldots, T.
$$
This leads to the simpler MDPDE  objective function, in analogue of (\ref{EQ:MDPDE_objFunc}), as given by 
\begin{eqnarray}
H_{n,\alpha}(\boldsymbol{\theta}) = \frac{1}{1+\alpha}\sum_{t=0}^T\sum_{i=1}^K \widehat{\pi}_{io}^{(n)}(t) \sum_{j=1}^K 
\left\{p_{ij}^{(n)}(t;\boldsymbol{\theta})^{1+\alpha} - \left(1 + \frac{1}{\alpha}\right)  
p_{ij}^{(n)}(t;\boldsymbol{\theta})^\alpha \widehat{\pi}_{ij}^{(n)}(t)\right\}.
\label{EQ:MDPDE_objFuncNS}
\end{eqnarray}
The resulting MDPDE can be studied asymptotically as $n\rightarrow\infty$ (fixed $T$) in a similar fashion as in Section \ref{SEC:Extn1},
provided we have a result similar to (\ref{EQ:CLT2}). 
Such results are available in the literature of Markov chain under suitable assumptions (e.g., \cite{Anderson/Goodman:1957})
which would lead to the asymptotic distribution of MDPDE as a suitable extension of (\ref{EQ:MDPDE_AN2}).
Considering the length and content of the present manuscript, we have deferred the detailed investigation of the properties
and applications of the MDPDE in such non-stationary context for a future work.


%
%

\section{Real-Life Application: Credit-Rating Migration Data}\label{SEC:real_data}

We now describe a real-life application of the finite Markov chain models in modelling credit-risk migration data. 
In a recent work, \cite{Mostel/etc:2020} statistically modeled Moody's corporate ratings migration data 
with finite Markov chain models for statistical inference. Here, we illustrate the usefulness of robust inference procedures 
for the finite Markov chain models in analyzing the average one year corporate transition rates as reported in Table 22 of \cite{Richhariya/etc:2001}.
They presented the average transition rates for seven rating categories AAA, AA, A, BBB, BB, B, CCC/C, along with the default (D),
computed based on the rating changes of different corporate entities till 2018 separately for three international markets, namely USA, Europe and
Emerging Markets; see pages 66-67 of \cite{Richhariya/etc:2001} for definitions 
(Sources: S\&P Global Fixed Income Research and S\&P Global Market Intelligence's CreditPro.).  
For USA, the data from the year 1981 are considered while for Europe and Emerging markets data from 1996 are used, 
taking into account the sample size issues.
The entities which were listed at the beginning of the time period but not listed at the final year of 2018, are reported in 
a  separate category NR (\textit{not-rated}).  
Our aim is to get parametric estimate of the upgrade and downgrade probabilities for each rating categories. 

We model these data, after conditioning by removal of the NR category, 
with a Markov chain model having $K=8$ states corresponding to the seven ratings and the default (D); 
these states are ordered from AAA to CCC/C with D being the lowest category (which is an absorbing state). 
We fit a parametric model with transition matrix $\boldsymbol{P}(\boldsymbol{\theta}) = (p_{ij}(\boldsymbol{\theta}))_{i,j=1, \ldots, K}$,
having a binomial-type form as in Example 2 (Section \ref{SEC:Ex}.2) with slight modifications. 
With an aim of predicting the probability of upgrade and downgrade from each states, 
we indeed club the transition probabilities to all states above and below the current state;
these combined data are presented in Table 1 of the Online Supplement (Section C), for all the three markets. 
Since the transition rate from any state to itself is the highest, the parametric model of Example 2 (Section \ref{SEC:Ex}.2) 
always yield an estimate of $\theta_i$ near about 0.5 for all $i$, which is clearly not suited to model our data. 
So, we consider a slightly modified model given by 
$$
p_{ii}(\boldsymbol{\theta}) = \theta_i^2, 
~~ \sum_{j<i} p_{ij}(\boldsymbol{\theta}) = (1-\theta_i)^2,
~~ \sum_{j>i} p_{ij}(\boldsymbol{\theta}) = 2\theta_i(1-\theta_i),
~~~~i=2, \ldots, K-1.
$$
The model for the first state (AAA) is considered as 
$p_{11}(\boldsymbol{\theta}) = \theta_1^2$ and $\sum_{j>1} p_{1j}(\boldsymbol{\theta}) = (1-\theta_1)$,
whereas the last state (D) is an absorbing state so that $p_{KK}(\boldsymbol{\theta})=1$ and $\sum_{j<K} p_{1j}(\boldsymbol{\theta}) = 0$.
Here, the parameter vector is $\boldsymbol{\theta}=(\theta_1, \ldots, \theta_{K-1})^t \in [0,1]^{K-1}$,
which we estimate from the given data (separately for the three markets) using the usual MLE 
and the proposed MDPDE with $\alpha=0.5$ (suggested choice of the tuning parameter).
Given an estimate $\widehat{{\theta}}_i$ of  $\theta_i$ for the $i$-th state ($i=2, \ldots, K-1$), 
the corresponding estimates of the upgrade and downgrade probabilities are obtained by $(1-\widehat{{\theta}}_i)^2$ 
and $2\widehat{{\theta}}_i(1-\widehat{{\theta}}_i)$, respectively. 
The resulting parameter estimates for all three markets are reported in Table \ref{TAB:Data_Est},
and the corresponding upgrade and downgrade probability estimates are presented (in \%) 
in Figure 1 in the Online Supplement (due to page restriction).

\begin{table}[h]
	\centering
	\caption{The MLE and the MDPDE at $\alpha=0.5$ of the parameters under the assumed Markov chain models 
		for the Credit rating migration data of three international markets}
	\begin{tabular}{ll|rr|rr|rr} \hline
Rating	&&	\multicolumn{2}{|c}{USA}			&	\multicolumn{2}{|c}{Europe}	&	\multicolumn{2}{|c}{Emerging Markets}		\\
State &&	MLE	&	MDPDE	&	MLE	&	MDPDE	&	MLE	&	MDPDE	\\\hline\hline
AAA	& ($\theta_1$) &		&		&		&		&		&		\\
AA	& ($\theta_2$)	&	0.9516	&	0.9540	&	0.9422	&	0.9418	&	0.9454	&	0.9524	\\
A	& ($\theta_3$)	&	0.9505	&	0.9638	&	0.9489	&	0.9629	&	0.9553	&	0.9683	\\
BBB	& ($\theta_4$)	&	0.9376	&	0.9668	&	0.9311	&	0.9676	&	0.9462	&	0.9653	\\
BB	& ($\theta_5$)	&	0.8959	&	0.9322	&	0.8937	&	0.9354	&	0.9184	&	0.9529	\\
B	& ($\theta_6$)	&	0.8988	&	0.9349	&	0.8897	&	0.9399	&	0.8895	&	0.9382	\\
CCC/C	& ($\theta_7$)	&	0.6803	&	0.7054	&	0.6666	&	0.6993	&	0.6563	&	0.7183	\\	
	\hline
	\end{tabular}
	\label{TAB:Data_Est}
\end{table}

It can be easily seen from the results that the MDPDE and the MLE are more close for the  higher-rating categories
and for developed markets; their difference increases for lower rating categories and also from USA to Europe followed by the emerging markets.
It is known that these ratings are more stable for the higher categories and more developed markets;
so, in these cases, there is no drastic effects of outlying observations and the proposed MDPDE yields similar parameter estimate as the MLE.
But, for the other group of lower rating category and developing markets, 
these ratings are known to be unstable due to sudden changes in economic conditions
and the MDPDE provides more stable results compared to the MLE which is quite far off in such cases. 
Similar patterns are also observed for the estimated upgrade and downgrade probabilities. 
Of particular interest is the downgrade probability from rating CCC/C, 
which is nothing but the default probability of a corporate entity after a year starting with a rating of CCC/C.
Based on the MLE, this default probability is highest for the emerging markets (45.11\%) followed by Europe (44.45\%) and USA (43.50\%),
whereas these estimates are 40.47\%, 42.06\% and 41.57\%, respectively, based on the proposed MDPDE.
It is then evident that the higher default probability of the emerging markets in the MLE based inference is only due 
to some outlying entities (sudden default due to external factors within these markets),  
although it indeed have a lowest (stable) default rate (starting from CCC/C) compared to Europe and USA
which is more intuitive and can be correctly obtained from the proposed MDPDE based inference yielding robust estimates.  
The same is also observed for the upgrade probability from CCC/C and for other lower ranking categories as well
(see Figure 1 in the Online Supplement, Section C).

\section{Concluding Remarks}

This paper develops a robust estimator, namely the minimum density power divergence estimator,
of the underlying parameter in finite Markov chain models and several important extensions.
The advantages of the proposed estimator is illustrated theoretically and empirically along 
with its applications in statistical hypotheses testing. 
Several possible extensions and future works are pointed out throughout the paper.
We would like to further mention that, although the simplest Wald-type tests based on the MDPDE are discussed here,
one might also want to construct robust versions of the other two types of tests, namely the likelihood-ratio and the score type tests.
Such constructions can also be done using the MDPDE and its asymptotic distributions (Theorems \ref{THM:MDPDE_conv} and \ref{THM:MDPDE_conv_Out}) 
under the present set-up of finite Markov chains, extending the corresponding works done for independent observations 
\citep{Basu/etc:2018,Basu/etc:2019} and will be investigated further in future.

Additionally, the present work opens up a new direction of research 
to further extend the concept of MDPDE for robust inference  in more complex stochastic processes. 
In particular, extensions for the Markov chains having countably infinite or continuous state-spaces, discrete time stochastic processes
and more generally continuous time stochastic processes would have significant advantages for robust insight generations 
in several real-life applications. Although the definition of the MDPDE can easily be extended for such general stochastic processes,
their theoretical study and practical implementations would need substantive additional (non-trivial) works.
We hope to pursue some of these important extensions in future.

\bigskip\noindent\textbf{Acknowledgment:}
The author wishes to thank the editor, the associate editor and two anonymous referees for 
their careful reading of the manuscript and several constructive suggestions to improve the paper.
This research is partially supported by the INSPIRE Faculty Research Grant
from  Department of Science and Technology, Government of India.



\begin{thebibliography}{3}
%
%

%

\bibitem[\protect\citeauthoryear{Anderson, T. W., and Goodman}{Anderson and Goodman}{1957}]{Anderson/Goodman:1957}
Anderson, T. W., and Goodman, L. A. (1957). 
Statistical inference about Markov chains. 
\textit{Ann Math Stat}, 89--110.


\bibitem[\protect\citeauthoryear{Basu, Basu and Jones}{Basu et~al.}{2006}]{Basu/etc:2006}
Basu, S., Basu, A. and Jones, M. C. (2006). Robust and efficient parametric estimation for censored survival data. 
{\em Ann Inst Stat Math}, { 58(2)}, 341--355.


\bibitem[\protect\citeauthoryear{Basu, Ghosh, Martin and Pardo}{Basu et~al.}{2018}]{Basu/etc:2018}
Basu, A., Ghosh, A., Martin, N. and Pardo, L. (2018)
Robust Wald-type tests for non-homogeneous observations based on minimum density power divergence estimator.
{\it Metrika}, 81, 493--522. 

\bibitem[\protect\citeauthoryear{Basu, Ghosh, Martin and Pardo}{Basu et~al.}{2019}]{Basu/etc:2019}
Basu, A., Ghosh, A., Martin, N. and Pardo, L. (2019)
A Robust Generalization of the Rao Test.
{\it ArXiv preprint}, 	arXiv:1908.09794 [stat.ME].


\bibitem[\protect\citeauthoryear{Basu, Harris, Hjort and Jones}{Basu et~al.}{1998}]{Basu/etc:1998}
Basu, A., Harris, I. R., Hjort, N. L., and Jones, M. C. (1998). 
Robust and efficient estimation by minimising a density power divergence. 
{\it Biometrika} {85}, 549--559.


\bibitem[\protect\citeauthoryear{Basu, Shioya and Park}{Basu et~al.}{2011}]{Basu/etc:2011}
Basu, A., Shioya, H. and Park, C. (2011).
\newblock {\em Statistical Inference: The Minimum Distance Approach}.
\newblock Chapman \& Hall/CRC. Boca Raton, Florida. 



\bibitem[\protect\citeauthoryear{Bertail et al.}{Bertail et al.}{2018}]{Bertail/etc:2018}
Bertail, P.; Ciolek, G. and Tillier, C. (2018). 
Robust Estimation for Markov Chains with Applications to Piecewise-deterministic Markov Processes.
\textit{Statistical Inference for Piecewise-deterministic Markov Processes}, Wiley Online Library, 107--146.


\bibitem[\protect\citeauthoryear{Billingsley}{Billingsley}{1961}]{Billingsley:1961}
Billingsley, P. (1961). 
Statistical methods in Markov chains. 
{\em Ann Math Stat}, 12--40.



\bibitem[\protect\citeauthoryear{Birge}{Birge}{1983}]{Birge:1983}
Birge, L. (1983). 
Robust testing for independent non identically distributed variables and Markov chains. 
\textit{Specifying Statistical Models}, Springer, 134--162.



\bibitem[\protect\citeauthoryear{Gani, J., and Jerwood}{Gani and Jerwood}{1971}]{Gani/Jerwood:1971}
Gani, J., and Jerwood, D. (1971). 
Markov chain methods in chain binomial epidemic models. 
\textit{Biometrics}, 27, 591--603.

%


\bibitem[\protect\citeauthoryear{Ghosh and Basu}{Ghosh and Basu}{2013}]{Ghosh/Basu:2013}
Ghosh, A. and Basu, A. (2013). 
Robust estimation for independent non-homogeneous observations 
using density power divergence with applications to linear regression. 
\textit{Electron. J Stat}, 7, 2420-2456.


%
%
%


\bibitem[\protect\citeauthoryear{Ghosh and Basu}{Ghosh and Basu}{2018}]{Ghosh/Basu:2018}
Ghosh, A., and Basu, A. (2018). 
Robust Bounded Influence Tests for Independent but Non-Homogeneous Observations.
{\it Statistica Sinica}, 28(3), 1133--1155.


\bibitem[\protect\citeauthoryear{Ghosh, Mandal, Martin and Pardo}{Ghosh et al.}{2016}]{Ghosh/etc:2016}
Ghosh, A., Mandal, A., Martin, N. and Pardo, L. (2016). 
Influence analysis of robust Wald-type tests.
{\it J Mult Anal}, \textbf{147}, 102--126.


\bibitem[\protect\citeauthoryear{Ghosh,  Martin, Basu and Pardo}{Ghosh et al.}{2018}]{Ghosh/etc:2018}
Ghosh, A., Martin, N., Basu, A., and Pardo, L. (2018). 
A new class of robust two-sample Wald-type tests. 
\textit{Int J Biostat}, 14(2).


\bibitem[\protect\citeauthoryear{Hampel, Ronchetti, Rousseeuw, and Stahel}{Hampel et~al.}{1986}]{Hampel/etc:1986}
Hampel, F. R., Ronchetti, E., Rousseeuw, P. J. and Stahel W.(1986). 
\textit{Robust Statistics: The Approach Based on Influence Functions}. 
New York, USA: John Wiley \& Sons. 

%


\bibitem[\protect\citeauthoryear{Hjort, N. L., and Varin}{Hjort and Varin}{2008}]{Hjort/Varin:2008}
Hjort, N.L., and Varin, C. (2008). 
ML, PL, QL in Markov chain models. 
\textit{Scand J Stat}, 35, 64-82.


%


\bibitem[\protect\citeauthoryear{Iosifescu}{Iosifescu}{2007}]{Iosifescu:2007}
Iosifescu, M. (2007). 
\textit{Finite Markov Processes and Their Applications. }
Dover Publications Inc., NY, USA.


\bibitem[\protect\citeauthoryear{Jones}{Jones}{2004}]{Jones:2004}
Jones, G. L. (2004). On the Markov chain central limit theorem. \textit{Prob Surveys}, 1(299-320), 5-1.



\bibitem[\protect\citeauthoryear{Kang and Lee}{Kang and Lee}{2014}]{Kang/Lee:2014}
Kang, J., and Lee, S. (2014). Minimum density power divergence estimator for Poisson autoregressive models. 
\textit{Comput Stat Data Anal}, 80, 44-56.

\bibitem[\protect\citeauthoryear{Kunsch}{Kunsch}{1984}]{Kunsch:1984}
Kunsch, H. (1984) 
Infinitesimal Robustness for Autoregressive Processes. 
\textit{Ann Stat}, 12(3), 843--863.


\bibitem[\protect\citeauthoryear{Lee and Lee}{Lee and Lee}{2010}]{Lee/Lee:2010}
Lee, S. and Lee, T. (2010). 
Robust estimation for order of hidden Markov models based on density power divergences. 
\textit{J Stat Comput Simul}, 80(5), 503--512,

\bibitem[\protect\citeauthoryear{Lee and Song}{Lee and Song}{2013}]{Lee/Song:2013}
Lee, S., and Song, J. (2013). Minimum density power divergence estimator for diffusion processes. 
\textit{Ann Inst Stat Math}, 65(2), 213--236.


\bibitem[\protect\citeauthoryear{Lifshits}{Lifshits}{1979}]{Lifshits:1979}
Lifshits, B. A. (1979). On the central limit theorem for Markov chains. 
\textit{Theory Prob Appl}, 23(2), 279--296.


\bibitem[\protect\citeauthoryear{Martin and Yohai}{Martin and Yohai}{1986}]{Martin/Yohai:1986}
Martin, R. D. and Yohai, V. J. (1986). 
Influence Functionals for Time Series. \textit{Ann Stat}, 1986, 14, 781--818.



\bibitem[\protect\citeauthoryear{Menendez et al.}{Menendez et al.}{1999}]{Menendez/etc:1999}
Menendez, M. L., Morales, D., Pardo, L., and Zografos, K. (1999). Statistical inference for finite Markov chains based on divergences. 
\textit{Stat Prob Letters}, 41(1), 9--17.

\bibitem[\protect\citeauthoryear{Menendez et al.}{Menrndez et al.}{2001}]{Menendez/etc:2001}
Menendez, M. L., Pardo, J. A., and Pardo, L. (2001). 
Csiszar's $\phi$-divergences for testing the order in a Markov chain. 
\textit{Stat Papers}, 42(3), 313--328.

\bibitem[\protect\citeauthoryear{Mostel et al.}{Mostel et al.}{2020}]{Mostel/etc:2020}
Mostel, L., Pfeuffer, M., and Fischer, M. (2020). 
Statistical inference for Markov chains with applications to credit risk. 
\textit{Comput Stat}, 35, 1659--1684.

\bibitem[\protect\citeauthoryear{Papapetrou and Kugiumtzis}{Papapetrou and Kugiumtzis}{2013}]{Papapetrou/Kugiumtzis:2013}
Papapetrou, M., and Kugiumtzis, D. (2013). 
Markov chain order estimation with conditional mutual information. 
\textit{Physica A}, 392(7), 1593--1601.


\bibitem[\protect\citeauthoryear{Rajarshi}{Rajarshi}{2014}]{Rajarshi:2014}
Rajarshi, M. B. (2014). 
\textit{Statistical inference for discrete time stochastic processes}. Springer. 


\bibitem[\protect\citeauthoryear{Richhariya et al.}{Richhariya et al.}{2019}]{Richhariya/etc:2001}
Richhariya, N. M., Jain, M., and Debnath, A. et al. (2019). 
Default, transition, and recovery: 2018 annual global corporate default and rating transition study. 
\textit{Standard \& Poor's Ratings Direct}. Available at {\tiny https://www.spratings.com/documents/20184/774196/2018AnnualGlobalCorporateDefaultAndRatingTransitionStudy.pdf}


\bibitem[\protect\citeauthoryear{Sirazhdinov, S. K., and Formanov}{Sirazhdinov and Formanov}{1984}]{Sirazhdinov/Formanov:1984}
Sirazhdinov, S. K., and Formanov, S. K. (1984). On estimates of the rate of convergence in the central limit theorem for homogeneous Markov chains. \textit{Theory Prob Appl}, 28(2), 229-239.




\bibitem[\protect\citeauthoryear{Zhao et al.}{Zhao et al.}{2001}]{Zhao/etc:2001}
Zhao, L. C., Dorea, C. C. Y., and Gonçalves, C. R. (2001). 
On determination of the order of a Markov chain. 
\textit{Stat Inf Stoch Proc}, 4(3), 273--282.

\end{thebibliography}
\end{document}


\title{Supplementary Material for \\``Robust Parametric Inference for Finite Markov Chains"
}

\author{Abhik Ghosh 
\\
Indian Statistical Institute, Kolkata, India  
\\
{\it abhik.ghosh@isical.ac.in
}}
\maketitle

%
%
%

\bigskip
\appendix
\section{Technical Derivations and Proofs}\label{SEC:intro}

\subsection{Derivation of Equation (4) of the main paper}
\label{SEC:MDPDE_Est}

From the formula of the likelihood $L_T(\boldsymbol{\theta})$, we get
\begin{eqnarray}
\log L_T(\boldsymbol{\theta}) &=& \sum_{i,j=1}^K \nu_{ij} \log p_{ij}(\boldsymbol{\theta})
\nonumber\\
&=& T \sum_{i,j=1}^K \widehat{\pi}_{io} \widehat{\pi}_{ij} \log p_{ij}(\boldsymbol{\theta})
\nonumber\\
&=& -  T \sum_{i,j=1}^K \widehat{\pi}_{io} \widehat{\pi}_{ij} \left[ \log\widehat{\pi}_{ij} - \log{p_{ij}(\boldsymbol{\theta})}\right] 
+ T \sum_{i,j=1}^K \widehat{\pi}_{io} \widehat{\pi}_{ij} \log\widehat{\pi}_{ij}
\nonumber\\
&=& -  T \sum_{i=1}^K \widehat{\pi}_{io} \sum_{j=1}^K \widehat{\pi}_{ij} \log \frac{\widehat{\pi}_{ij}}{p_{ij}(\boldsymbol{\theta})} 
+ T \sum_{i=1}^K \widehat{\pi}_{io} \sum_{j=1}^K \widehat{\pi}_{ij} \log\widehat{\pi}_{ij},
\nonumber
\end{eqnarray}
which lead to Equation (4) of the main paper.

\subsection{Proof of Theorem 2.2}

We prove Theorem 2.2 (of the main paper) by using an argument similar to the one used by \cite{Menendez/etc:1999} for disparities. 
Note that $\Im_C \subset \mathcal{L}_C$, the interior of the $c$-dimensional unit cube. 
Consider a neighborhood $V$ of $\boldsymbol{\theta}_\pi$ such that $\boldsymbol{P}(\boldsymbol{\theta})$ 
has continuous partial derivatives for all $\boldsymbol{\theta}\in V \subseteq\Theta$;
this is possible in view of Assumption (A2). 
Then, with slight abuse of notation, we consider the function 
$$
\boldsymbol{U}_{\alpha}(\boldsymbol{\Pi}_C, \boldsymbol{\theta})
= \boldsymbol{U}_{\alpha}(\boldsymbol{\Pi}_C, \boldsymbol{P}_C(\boldsymbol{\theta})) : 
\mathcal{L}_C\times V \mapsto \mathbb{R}^d
$$ 
where each coordinate function is continuous in $\boldsymbol{\theta}\in V$. 
By definition, for $\boldsymbol{\Pi}=\boldsymbol{\Pi}^o$, we have 
$\boldsymbol{U}_{\alpha}(\boldsymbol{\Pi}^o_C, \boldsymbol{\theta}_\pi)
= \boldsymbol{U}_{\alpha}(\boldsymbol{\Pi}^o_C, \boldsymbol{\Pi}^o_C) = \boldsymbol{0}_d$,
i.e., the function $\boldsymbol{U}_\alpha$ has a zero at 
$(\boldsymbol{\Pi}_C, \boldsymbol{\theta}) = (\boldsymbol{\Pi}^o_C, \boldsymbol{\theta}_\pi)$.

Next, through standard differentiation, we get 
\begin{eqnarray}
\frac{\partial}{\partial\boldsymbol{\theta}}\boldsymbol{U}_{\alpha}(\boldsymbol{\Pi}^o_C, \boldsymbol{\theta}_\pi)
=\boldsymbol{\Psi}_\alpha(\boldsymbol{\Pi}^o, \boldsymbol{\theta}_\pi)
~~~\mbox{ and }~~
\frac{\partial}{\partial\boldsymbol{\Pi}_C}\boldsymbol{U}_{\alpha}(\boldsymbol{\Pi}^o_C, \boldsymbol{\theta}_\pi)
=\boldsymbol{J}(\boldsymbol{\theta}_\pi)^t\boldsymbol{B}_\alpha(\boldsymbol{\Pi}^o, \boldsymbol{\theta}_\pi)^{-1}
\label{EQ:diff}
\end{eqnarray}
Since $\boldsymbol{\Psi}_\alpha(\boldsymbol{\Pi}^o, \boldsymbol{\theta}_\pi)$ is non-singular by Assumption (A3), 
we can now apply implicit function theorem on the function 
$\boldsymbol{U}_{\alpha}(\boldsymbol{\Pi}_C, \boldsymbol{\theta})$ 
at the point $(\boldsymbol{\Pi}_C, \boldsymbol{\theta}) = (\boldsymbol{\Pi}^o_C, \boldsymbol{\theta}_\pi)$
to get a neighborhood $W$ of $\boldsymbol{\Pi}_C^o$ in $\mathcal{L}_C$ and a unique continuously differentiable function
$\widetilde{\boldsymbol{\theta}}: W \mapsto  \mathbb{R}^d$ such that  
$\widetilde{\boldsymbol{\theta}}(\boldsymbol{\Pi}^o_C) = \boldsymbol{\theta}_\pi.$ and
\begin{eqnarray}
&&\boldsymbol{U}_{\alpha}(\boldsymbol{\Pi}_C, \widetilde{\boldsymbol{\theta}}(\boldsymbol{\Pi}_C)) = \boldsymbol{0}_d, 
~~~~\mbox{for all } ~~ \boldsymbol{\Pi}_C \in W.
\nonumber
\end{eqnarray}
Differentiating this last equation with respect to $\boldsymbol{\Pi}_C$,  via Chain rule, we get
\begin{eqnarray}
\frac{\partial \boldsymbol{U}_{\alpha}(\boldsymbol{\Pi}_C, \widetilde{\boldsymbol{\theta}}(\boldsymbol{\Pi}_C)) 
}{\partial\boldsymbol{\Pi}_C}
+ \frac{\partial\boldsymbol{U}_{\alpha}(\boldsymbol{\Pi}_C, \widetilde{\boldsymbol{\theta}}(\boldsymbol{\Pi}_C)) 
}{\partial\widetilde{\boldsymbol{\theta}}(\boldsymbol{\Pi}_C)}
\frac{\partial\widetilde{\boldsymbol{\theta}}(\boldsymbol{\Pi}_C)}{\partial\boldsymbol{\Pi}_C}
= \boldsymbol{0}_d, 
~~~~\mbox{for all } ~~ \boldsymbol{\Pi}_C \in W.
\nonumber
\end{eqnarray}
Evaluating it at $\boldsymbol{\Pi}_C = \boldsymbol{\Pi}^o_C$ and simplifying using (\ref{EQ:diff}), 
we get 
\begin{eqnarray}
\left.\frac{\partial \widetilde{\boldsymbol{\theta}}(\boldsymbol{\Pi}_C)}{\partial\boldsymbol{\Pi}_C}
\right|_{\boldsymbol{\Pi}_C = \boldsymbol{\Pi}^o_C}
=\boldsymbol{\Psi}_\alpha(\boldsymbol{\Pi}^o, \boldsymbol{\theta}_\pi)^{-1}
\boldsymbol{J}(\boldsymbol{\theta}_\pi)^t\boldsymbol{B}_\alpha(\boldsymbol{\Pi}^o, \boldsymbol{\theta}_\pi)^{-1}.
\nonumber
\end{eqnarray}  
But, a Taylor series expansion of $ \widetilde{\boldsymbol{\theta}}(\boldsymbol{\Pi}_C)$ 
around $\boldsymbol{\Pi}^o_C$ yields
\begin{eqnarray}
\widetilde{\boldsymbol{\theta}}(\boldsymbol{\Pi}_C) = \widetilde{\boldsymbol{\theta}}(\boldsymbol{\Pi}^o_C)
+ \left.\frac{\partial \widetilde{\boldsymbol{\theta}}(\boldsymbol{\Pi}_C)}{\partial\boldsymbol{\Pi}_C}
\right|_{\boldsymbol{\Pi}_C = \boldsymbol{\Pi}^o_C} \left(\boldsymbol{\Pi}_C - \boldsymbol{\Pi}^o_C\right)
+ o\left(||\boldsymbol{\Pi}_C - \boldsymbol{\Pi}^o_C||\right).
\nonumber
\end{eqnarray}
Therefor, upon simplification, for any $\boldsymbol{\Pi}_C \in W$, we get
\begin{eqnarray}
\widetilde{\boldsymbol{\theta}}(\boldsymbol{\Pi}_C) - \boldsymbol{\theta}_\pi
=\boldsymbol{\Psi}_\alpha(\boldsymbol{\Pi}^o, \boldsymbol{\theta}_\pi)^{-1}
\boldsymbol{J}(\boldsymbol{\theta}_\pi)^t\boldsymbol{B}_\alpha(\boldsymbol{\Pi}^o, \boldsymbol{\theta}_\pi)^{-1}\left(\boldsymbol{\Pi}_C - \boldsymbol{\Pi}^o_C\right)
+ o\left(||\boldsymbol{\Pi}_C - \boldsymbol{\Pi}^o_C||\right).
\label{EQ:thm_Eq1}
\end{eqnarray}

Finally, in view of the result (8) of the main paper, we have $\widehat{\boldsymbol{\Pi}}_C \rightarrow \boldsymbol{\Pi}^o_C$ 
almost surely and $\sqrt{T}\left(\widehat{\boldsymbol{\Pi}}_C - \boldsymbol{\Pi}^o_C\right)=O_P(1)$ as $T\rightarrow \infty$.
Thus $\widehat{\boldsymbol{\Pi}}_C\in W$ almost surely for sufficiently large $T$ 
and hence $\widetilde{\boldsymbol{\theta}}(\widehat{\boldsymbol{\Pi}}_C)$ is the unique solution of the equations
$$
\boldsymbol{U}_{\alpha}(\widehat{\boldsymbol{\Pi}}_C, \boldsymbol{P}_C(\boldsymbol{\theta})) = \boldsymbol{0}_d,
~~~~\mbox{ or equivalently, }~~ 
\boldsymbol{U}_{\alpha}(\widehat{\boldsymbol{\Pi}}, \boldsymbol{P}(\boldsymbol{\theta}))
= \boldsymbol{U}_{T,\alpha}(\boldsymbol{\theta}) = \boldsymbol{0}_d,
$$
which is the MDPDE estimating equation. 
Therefore, $\widetilde{\boldsymbol{\theta}}(\widehat{\boldsymbol{\Pi}}_C)$ is indeed our target MDPDE 
$\widehat{\boldsymbol{\theta}}_\alpha$ and also almost surely unique.
We can verify that it satisfies the required relation in Eq.~(12) of the main paper
by substituting $\boldsymbol{\Pi}_C = \widehat{\boldsymbol{\Pi}}_C$ in Equation (\ref{EQ:thm_Eq1})
completing the proof of Part (i) of the theorem.

\bigskip\noindent
Part (ii) of the theorem follows directly from the relation (12)  
and the result (8) 
of the main paper.

\hfill{$\square$}

\subsection{Proof of Theorem 2.3}

We start with the equation 
$\boldsymbol{U}_{\alpha}(\boldsymbol{\Pi}_\epsilon, \boldsymbol{P}(\boldsymbol{F}_\alpha(\boldsymbol{\Pi}_\epsilon))) =\boldsymbol{0}_d$, 
and differentiate it with respect to $\epsilon$ to get (using Chain rule of differentiation)
$$
\left. \frac{\partial\boldsymbol{U}_{\alpha}(\boldsymbol{\Pi}, \boldsymbol{P}(\boldsymbol{F}_\alpha(\boldsymbol{\Pi}_\epsilon)))}{
\partial\boldsymbol{\Pi}}\right|_{\boldsymbol{\Pi} = \boldsymbol{\Pi}_\epsilon} \frac{\partial \boldsymbol{\Pi}_\epsilon}{\partial\epsilon}
+ \left. \frac{\partial \boldsymbol{U}_{\alpha}(\boldsymbol{\Pi}_\epsilon, \boldsymbol{P}(\boldsymbol{\theta}))}{
\partial\boldsymbol{\theta}}\right|_{\boldsymbol{\theta} = \boldsymbol{F}_\alpha(\boldsymbol{\Pi}_\epsilon)} 
\frac{\partial \boldsymbol{F}_\alpha(\boldsymbol{\Pi}_\epsilon)}{\partial \epsilon}
= \boldsymbol{0}_d.
$$
Now, substituting $\epsilon=0$ in the above equation, and using the definition of the influence function and the matrix 
$\boldsymbol{\Psi}_\alpha(\boldsymbol{\Pi}, \boldsymbol{\theta})$, we get 
$$
\sum_{i=1}^K {\pi}_{io}  \left[-\sum_{j=1}^K \boldsymbol{\psi}_{ij}(\boldsymbol{\theta}_\pi)p_{ij}(\boldsymbol{\theta}_\pi)^{\alpha}{\pi}_{ij}^o  
+ \boldsymbol{\psi}_{it_i}(\boldsymbol{\theta}_\pi)p_{it_i}(\boldsymbol{\theta}_\pi)^{\alpha}\right]
+ \boldsymbol{\Psi}_\alpha(\boldsymbol{\Pi}^o, \boldsymbol{\theta}_\pi)	IF(\boldsymbol{t}; \boldsymbol{F}_\alpha, \boldsymbol{\Pi}^o) 
= \boldsymbol{0}_d.
$$
After simplification, it leads to the form of the influence function as given in Eq.~(13) of the main paper, completing  the proof of the theorem. 
\hfill{$\square$}

\bigskip
\section{Another Interesting Example }
\label{SEC:Ex}


Here we present an interesting example, where the MDPDE equals to the MLE for each $\alpha\geq 0$. 
Consider the simple random walk with reflecting barriers having state-space $S=\{1, 2, \ldots, K \}$ and a parametric transition matrix as given by
\begin{eqnarray}
\boldsymbol{P}(\theta) = \begin{bmatrix}
\begin{array}{ccccccc}
0 		 & 1 		& 0 		& 0 	 & \cdots & 0  & 0\\
1-\theta & 0 		& \theta 	& 0 	 & \cdots & 0  & 0\\
0 		 & 1-\theta & 0 		& \theta & \cdots & 0  & 0\\
: 		 & 		: 	& 	: 		&  :	 &  \ddots & :  & : \\
0 		 &    0		&  	0		&  0     & \cdots  & 1  & 0\\
\end{array}
\end{bmatrix}.
\label{EX1:P}
\end{eqnarray}
Here our target parameter $\theta$ is a scalar and the associated parameter space is $\Theta = [0,1]$.
It is easy to verify that this Markov chain is stationary and ergodic with initial (stationary) probabilities being 
$$
\pi_{io}=p_{io}(\theta) = \pi_{1o} \theta^{i-2} (1-\theta)^{1-i}, ~~~i=2, \ldots, K-1;  
~~~\pi_{Ko}=p_{Ko}(\theta) = \pi_{1o} \theta^{K-2} (1-\theta)^{2-K},
$$
where $\pi_{1o}=p_{1o}(\theta)$ is defined from the relation $\sum_{i=1}^K \pi_{io} =1$.
Further, Assumptions (A1)--(A3) hold for $\boldsymbol{P}(\theta)$ in (\ref{EX1:P})
with $C=\{ (1,2); (i, i+1), (i, i-1) \mbox{ for } i=2, 3, \ldots, K-1; (K, K-1) \}$ and, 
hence, $c=2(K-1)$ and $\boldsymbol{J}(\theta) = (0, 1, -1, 1, -1, \ldots, 1, -1, 0)^t$. 

Let us now consider the problem of estimating $\theta$ from an observed sequence 
$\mathcal{X}_T = \{X_0, X_1, \ldots, X_T\}$ form this Markov chain model. Firstly, the MLE of $\theta$ is given by 
$$
\widehat{\theta}_0 = \frac{\sum_{i=2}^{K-1} \nu_{i(i+1)}}{\sum_{i=2}^{K-1} [\nu_{i(i-1)}+\nu_{i(i+1)}]} 
= \frac{\sum_{i=2}^{K-1} \nu_{i(i+1)}}{\sum_{i=2}^{K-1} \nu_{i+}}
= \frac{\sum_{i=2}^{K-1} \widehat{\pi}_{io} \widehat{\pi}_{i(i+1)}}{\sum_{i=2}^{K-1} \widehat{\pi}_{io}}.
$$
Next, to find the MDPDE of $\theta$ with tuning parameter $\alpha\geq 0$, 
we simplify the corresponding  estimating equation 
[Eq.~(6) of the main paper] as
\begin{eqnarray}
\sum_{i=2}^{K-1} \left\{ \nu_{i+} \left[\theta^\alpha - (1-\theta)^\alpha\right] 
- \left[\nu_{i(i+1)} \theta^{\alpha-1} - \nu_{i(i-1)} (1-\theta)^{\alpha-1}\right]\right\} = 0. 
\label{EQ:Ex1_MDPDE_EstEq}
\end{eqnarray}
Although the above estimating equation (\ref{EQ:Ex1_MDPDE_EstEq}) is not directly solvable analytically, 
one can easily verify that the MLE $\widehat{\theta}_0$ is indeed a solution of (\ref{EQ:Ex1_MDPDE_EstEq}) for any $\alpha\geq 0$.
Therefore, the MDPDEs for all $\alpha\geq 0$ are the same, given by $\widehat{\theta}_0$, for this example.
Additionally, since (A1)--(A3) hold, one can obtain its asymptotic properties at the model from Theorem 2.1 of the main paper.
In particular, with some algebra,  we have 
\begin{eqnarray}
\boldsymbol{\Psi}_\alpha(\boldsymbol{P}(\theta), {\theta}) 
&=& [1-p_{1o}(\theta) - p_{Ko}(\theta))]\left[(1-\theta)^{\alpha-1} + \theta^{\alpha-1}\right],
\nonumber\\
\boldsymbol{\Omega}_\alpha(\boldsymbol{P}(\theta), {\theta}) &=& [1-p_{1o}(\theta) - p_{Ko}(\theta))]\theta(1-\theta)
\left[(1-\theta)^{\alpha-1} + \theta^{\alpha-1}\right]^2.
\nonumber
\end{eqnarray}
Thus, although these two quantities depend on $\alpha$, the asymptotic variance of the MDPDE becomes independent of $\alpha$
and is given by $\theta(1-\theta)/[1-p_{1o}(\theta) - p_{Ko}(\theta))]$.
This is consistent with the fact that the MDPDEs themselves do not depend on $\alpha$.
Further, the above asymptotic variance formula is exactly the same as derived in \cite{Hjort/Varin:2008} for the MLE
and thus our Theorem 2.1 generalizes their results for the larger class of MDPDEs. 

We conjecture that the MDPDEs will be independent of $\alpha$ and hence coincide with the MLE having no robustness benefit,
as in the present example, whenever the transition matrix $\boldsymbol{P}(\theta)$ has elements as a linear function 
of parameters only. This is certainly an interesting phenomenon 
which was never observed so prominently in the literature of DPD and its wide range of applications.

\section{Additional Table and Figures for the Real Data Application}

\begin{table}[h]
	\centering
	\caption{Combined (average) one year corporate transition rates (in \%) for downgrade, upgrade and remaining steady, 
		starting from any rating state at the beginning of the year, for the 2018 Credit rating migration data of three international markets}
\begin{tabular}{l|rrr|rrr|rrr} \hline
Rating	at &	\multicolumn{3}{|c}{USA}			&	\multicolumn{3}{|c}{Europe}	&	\multicolumn{3}{|c}{Emerging Markets}		\\
the Start &	Down	&	Up	&	Steady	&	Down	&	Up	&	Steady	&	Down	&	Up	&	Steady	\\\hline\hline
AAA	&	9.79	&	--	&	90.21	&	12.73	&	--	&	87.27	&	9.63	&	--	&	90.37	\\
AA	&	8.61	&	0.53	&	90.86	&	11.00	&	0.28	&	88.72	&	8.61	&	1.16	&	90.24	\\
A	&	6.22	&	1.84	&	91.94	&	6.33	&	1.94	&	91.73	&	5.44	&	1.75	&	92.81	\\
BBB	&	4.82	&	3.83	&	91.34	&	4.32	&	4.73	&	90.95	&	5.64	&	2.56	&	91.80	\\
BB	&	9.66	&	5.58	&	84.76	&	8.76	&	6.25	&	84.99	&	6.69	&	4.81	&	88.50	\\
B	&	9.28	&	5.48	&	85.24	&	7.47	&	7.30	&	85.23	&	7.81	&	7.15	&	85.04	\\
CCC/C	&	33.71	&	15.11	&	51.18	&	31.42	&	17.63	&	50.95	&	22.41	&	23.16	&	54.43	\\
\hline
\end{tabular}
\label{TAB:Data_Est}
\end{table} 

\begin{figure}[h]
	\centering
	\subfloat[USA]{
		\includegraphics[width=0.45\textwidth]{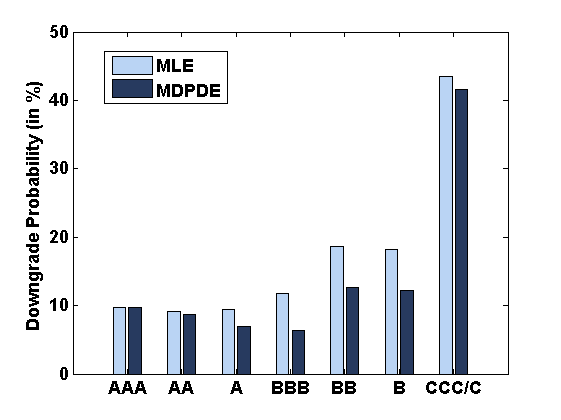}
		\includegraphics[width=0.45\textwidth]{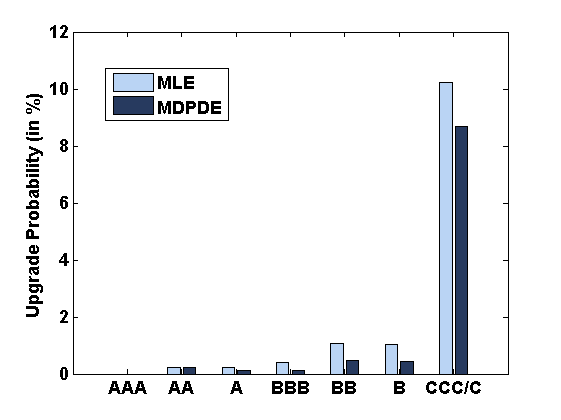}
		\label{FIG:Data_USA_DowngradeP}}
\\
	\subfloat[Europe]{
		\includegraphics[width=0.45\textwidth]{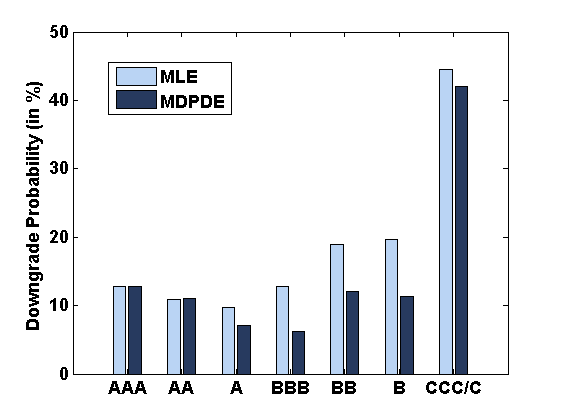}
		\includegraphics[width=0.45\textwidth]{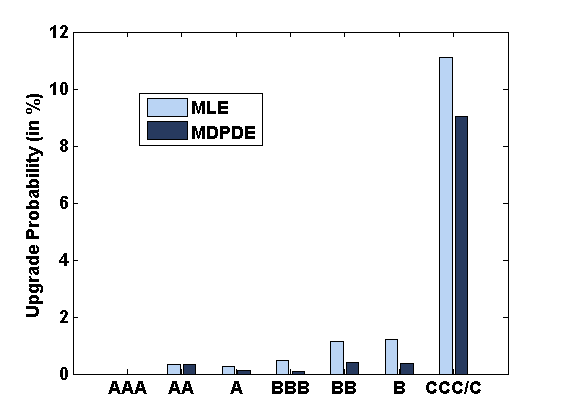}
		\label{FIG:Data_USA_DowngradeP}}
\\
	\subfloat[Emerging Markets]{
	\includegraphics[width=0.45\textwidth]{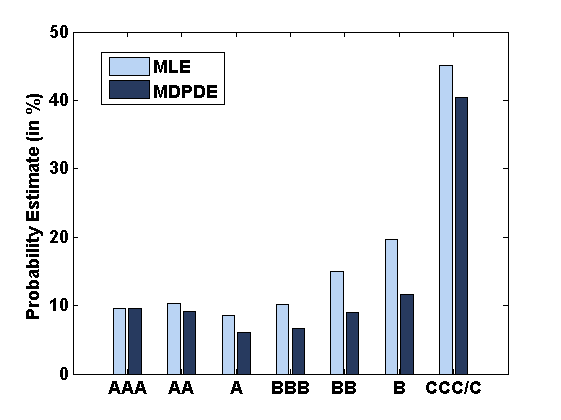}
	\includegraphics[width=0.45\textwidth]{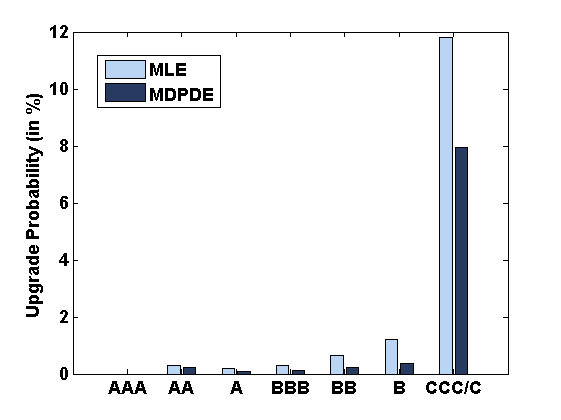}
	\label{FIG:Data_USA_DowngradeP}}
\caption{Estimated downgrade and upgrade probabilities (in \%), obtained via the MLE and the MDPDE at $\alpha=0.5$, 
	for the Credit rating migration data of three international markets}
\label{FIG:Ex4_MSE}
\end{figure}

\clearpage